\documentclass[iop,onecolumn]{emulateapj}

\usepackage{natbib}
\catcode`\@=11
\newcommand{\gapprox}{\mathrel{\mathpalette\@versim>}}
\newcommand{\lapprox}{\mathrel{\mathpalette\@versim<}}
\newcommand{\propapprox}{\mathrel{\mathpalette\@versim\propto}}
\newcommand{\@versim}[2]
  {\lower3.1truept\vbox{\baselineskip0pt\lineskip0.5truept
\ialign{$\m@th#1\hfil##\hfil$\crcr#2\crcr\sim\crcr}}}
\catcode`\@=12

\shorttitle{X-rays and gamma rays from cavity SNRs}

\begin{document}

\title{X-ray and gamma-ray emission from middle-aged supernova remnants in
cavities:  1.  Spherical symmetry}

\author{Zhu Tang,\altaffilmark{1}
Stephen P. Reynolds,\altaffilmark{1}
\&\ Sean M. Ressler\altaffilmark{1,2}
}

\altaffiltext{1}{Department of Physics, North Carolina State University, 
Raleigh, NC 27695-8202; ztang2@unity.ncsu.edu} 
\altaffiltext{2}{Physics Department, University of California, Berkeley,
Berkeley, CA 94720}

\begin{abstract}

We present analytical and numerical studies of models of
supernova-remnant (SNR) blast waves expanding into uniform media and
interacting with a denser cavity wall, in one spatial dimension.  We
predict the nonthermal emission from such blast waves: synchrotron
emission at radio and X-ray energies, 
and bremsstrahlung, inverse-Compton
emission (from cosmic-microwave-background seed photons; ICCMB), and
emission from the decay of $\pi^0$ mesons produced in inelastic
collisions between accelerated ions and thermal gas, at GeV and TeV
energies.  Accelerated particle spectra are assumed to be power-laws
with exponential cutoffs at energies limited by the remnant age or
(for electrons, if lower) by radiative losses.  We compare the results
with those from homogeneous (``one-zone'') models.  Such models give
fair representations of the 1-D results for uniform media, but
cavity-wall interactions produce effects for which one-zone models are
inadequate.  We study the time evolution of SNR morphology and
emission with time.  Strong morphological differences exist between
ICCMB and $\pi^0$-decay emission; at some stages, the TeV emission can
be dominated by the former and the GeV by the latter, resulting in
strong energy-dependence of morphology.  Integrated gamma-ray spectra
show apparent power-laws of slopes that vary with time, but do not
indicate the energy distribution of a single population of particles.
As observational capabilities at GeV and TeV energies improve, spatial
inhomogeneity in SNRs will need to be accounted for.

\end{abstract}

\keywords{
ISM: supernova remnants ---
X-rays: ISM 
}

\section{Introduction}
\label{intro}

Supernova remnants, along with pulsar-wind nebulae, are the most
obvious examples of Galactic fast-particle production.  GeV electrons
have been known to be present since the widespread adoption of
Shklovsky's 1953 suggestion of synchrotron emission as the process
producing radio emission from SNRs \citep{shklovsky53}. The range of
observable electron energies was extended to the TeV range with the
discovery of X-ray synchrotron emission from SN 1006
\citep{rc81,koyama95}.  A dozen or so Galactic shell supernova
remnants (SNRs) now show evidence for X-ray synchrotron emission at
least in localized areas.  See \cite{reynolds08}, R08, for a review.

The advent of GeV/TeV astronomy has made possible the more detailed
characterization of relativistic-particle populations in SNRs.  Of the
remnants showing X-ray synchrotron emission, TeV emission with a shell
morphology is observed from G347.3-0.5 (RX J1713.7-3946), SN 1006, RCW
86, G266.2-1.2 (``Vela Jr.''), and HESS J1731-347 (see \cite{rieger13}
for a review), along with unresolved or complex emission from
Cassiopeia A and Tycho \citep{acciari10,giordano12}.  This emission
could be due to leptonic processes (bremsstrahlung or inverse-Compton
[IC] emission) or hadronic processes (inelastic pp scattering
producing pions, of which $\pi^0$'s decay to photons) (see R08 for a
review).  Firm evidence for hadronic emission would demonstrate
directly that SNRs produce cosmic-ray ions, and would be an important
advance.  While various authors \citep[e.g.,][]{morlino12,berezhko08}
suggest that this is the case in several of the above objects, the
one-zone models typically used are sufficiently oversimplified that
there is still room for debate (e.g., \cite{atoyan12} for Tycho).  The
{\sl Fermi} LAT instrument has added considerably to the observational
constraints that models must meet; GeV observations of G347.3-0.5
\citep{abdo11} appear to disfavor hadronic models
\citep[e.g.,][]{ellison12}, while those of Cas A can support a hybrid
model involving both mechanisms \citep[e.g.,][]{araya10}.

{\sl Fermi} has also demonstrated the existence of a largely
unanticipated class of SNR sources, older objects interacting with
molecular clouds \citep[e.g.,][]{abdo10}. In these cases, X-ray
synchrotron emission is absent, as the shock waves are currently too
slow to accelerate electrons to the required TeV energies.  However,
GeV protons can easily produce GeV photons through the $\pi^0$
process, and these objects have been successfully modeled this way
\citep[e.g., IC 443 and W44][]{ackermann13}.  Even here, though, it is
not absolutely clear if newly accelerated protons are necessary;
simple compression of ambient cosmic-ray protons may be sufficient
\citep{uchiyama10}.

\begin{figure}
\centerline{\includegraphics[scale=2.2]{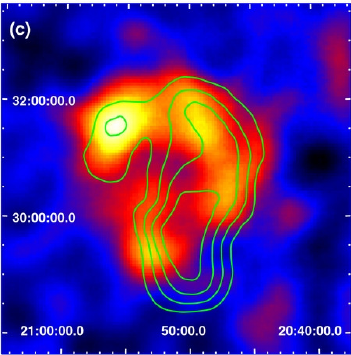}}
\caption{\footnotesize {\sl Fermi} image of the Cygnus Loop, with
radio contours \citep{katagiri11}.
}
\label{cygloop}
\vskip-0.1truein
\end{figure}

A potential third class of {\sl Fermi} SNR sources may exist:
middle-aged ($t \gapprox 2000$ yr) SNRs which are interacting with
low-density media \citep[e.g., Pup A;][]{hewitt12}.  The Cygnus Loop
(Fig.~\ref{cygloop}) could belong to this class.
%the proposed prototype. along with Pup A and S147.  
Here the ambient or internal gas densities may not be sufficient for a
hadronic explanation, especially since the Cygnus Loop appears to have
resulted from an explosion in a wind-blown cavity.  In this paper we
explore the GeV/TeV emission to be expected from a SNR exploding in a
low-density cavity, subsequently interacting with the cavity wall.

Cavity explosions have been the subject of theoretical investigation
for many years.  Early 1-D numerical hydrodynamic studies were done
by \cite{tenorio90,tenorio91} focusing on realistic stellar-wind
environments, and on SNRs that had reached the radiative phase.  These
studies predicted optical and thermal X-ray emission from such
interactions.  \cite{dwarkadas05} examined this problem, focusing
on earlier stages of evolution, and on thermal X-ray emission.

Simulations of the interaction of a fast Wolf-Rayet (WR) wind with a
slower, previous red supergiant (RSG) wind show that the shocked fast
wind can occupy the bulk of the cavity volume, with material of
roughly constant density, bounded by a thin shell with very much
higher density \citep{vanmarle12}.  Such cavities can be as large as
10 pc or greater \citep{vanmarle15}.  Since we are interested in later
stages of SNR evolution in which any unshocked stellar wind has
already been swept up, and for which the SNR blast wave is in the
Sedov phase, we shall assume that the remaining cavity with which the
blast wave interacts has constant density.  We shall also assume that
the shell bounding the cavity is sufficiently thick that the blast
wave has not emerged by the time our simulations end.  Unlike most
previous studies, we focus on nonthermal emission, in particular that
produced just before and for some time after the interaction.

A few calculations of SNR evolution have been published using
numerical hydrodynamics coupled with realistic shock-acceleration
physics, predicting both thermal and nonthermal emission
\citep[e.g.,][]{lee13}.  However, such models are computationally
expensive and have been reported for only a few cases.  Most GeV/TeV
modeling of SNRs has been done in a ``one-zone'' (or zero-dimensional)
context: a spatially homogeneous source (although often with
considerable sophistication in the spectral modeling , e.g.,
\cite{abdo09} among many others).  An important question is the extent
to which such modeling reflects realistic sources.  Here we begin an
investigation of this question with a suite of spherically symmetric
(1-D) models of cavity explosions. These models are compared with
one-zone and analytic models, and with 1-D models of expansion into a
uniform medium.  A larger range of models, with parameter studies, is
given in Tang (2016; PhD thesis, in preparation).

\section{Dynamics, Particle Acceleration, and Magnetic-Field Evolution}

We consider two cases: evolution in a uniform medium, and evolution in
a cavity of uniform density ($n_0 = 0.25$ cm$^{-3}$) bounded by a wall
(at $R = 2.67 \times 10^{19}$ cm) of much higher but still uniform
density (20 times higher).  In the uniform case, we assume all ejecta
have been reheated by the reverse shock, corresponding to the Sedov
self-similar solution.  In the cavity case, the interaction of the
blast wave with the cavity wall drastically slows the blast wave while
creating a reflected shock wave that moves back toward the remnant
center.  In one dimension, the reflected shock is eventually reflected
back out and returns to the blast wave.  However, we do not include
cooling in our simulations, so we consider only shock velocities above
about 300 km s$^{-1}$ (see below), and for the parameter ranges we
consider, by the time the blast wave speed has dropped to this value,
the reflected shock has not reached the center yet.  We treat the
uniform case both analytically and numerically; the cavity case is
treated numerically only.  We use the versatile grid-based
hydrodynamics code VH-1, a conservative, finite-volume code for
solving the equations of ideal hydrodynamics, based on the Lagrangian
remap version of the Piecewise Parabolic Method \citep{colella84}.
The code is third order in space and second order in time.  We assume
that the blast wave dynamics are not affected by cosmic-ray pressure
(the low-efficiency limit).  The grid of 1000 linearly spaced radial
zones expands to follow the moving blast wave.

We neglect cooling in these simulation.  \cite{blondin98} give the
characteristic time for the cooling transition for a blast wave in 
a uniform medium as
\begin{equation}
t_{\rm tr} = 2.9 \times 10^4\,E_{51}^{4/17}\,n_0^{-9/17} \ {\rm yr}
\end{equation}
where $E_{51}$ is the explosion energy in units of $10^{51}$ erg,
and $n_0$ is the upstream number density (assuming cosmic abundances).
At this time, the blast-wave velocity is
\begin{equation}
v_{\rm shock}(t_{\rm tr}) = 260\, E_{51}^{1/17} \, n_0^{2/17}.
\end{equation}
We use the second criterion, since for a cavity explosion, the blast
wave will reach the cavity wall rapidly, then decelerate fairly
quickly.  The material in the cavity will have a much longer cooling
time.  We follow our simulations to blast-wave speeds of about 320 km
s$^{-1}$, slightly higher than $v_{\rm shock}(t_{\rm tr}) \sim 310$ km s$^{-1}$
for an upstream density $n_0 = 5$ cm$^{-3}$ and $E_{51} = 1$.  

\subsection{Particle acceleration}

Relativistic particles are assumed accelerated at the shock front to a
maximum energy limited by either the finite remnant age or (for
electrons) radiative losses, if that limitation is more restrictive.
A constant fraction $\zeta = 10^{-4}$ of ions with momentum above some
value (here taken to be ten times the thermal proton momentum $m_p
v_{\rm shock}$) is assumed to be accelerated, with a power-law
distribution of specified slope.  The same spectrum of electrons, with
a normalization lower by a factor $k_{ei} = 0.02$, is assumed to be
produced.  (These values are typical of shock-acceleration models;
see, e.g., \cite{lee13}).  Maximum energies are calculated for each
fluid element, and the particle population then evolves by adiabatic
(and for electrons, synchrotron) losses.  Diffusion is neglected.
This treatment of the particle populations is similar to that used in
\cite{reynolds98}, R98, differing slightly only in now normalizing the
accelerated-particle distributions by $\zeta$ and $k_{ei}$.

We consider accelerated ions to have a simple power-law momentum
distribution, $N_i({\bf p}) = K_p {\bf p}^{-\sigma}$.  In standard
test-particle DSA, $\sigma = 3r/(r - 1)$ where $r$ is the shock
compression ratio.  Then for the
magnitude of momentum, $N_i(p)dp = (4 \pi p^2 dp) N_i({\bf p}) dp
= 4\pi K_p\, p^{2 - \sigma}\, dp$.  Then the energy distribution is
\begin{equation}
N_i(E) = N_i(p) {dp \over dE} \equiv K_i E^{-b} \ \Rightarrow \ 
   K_i = 4\pi K_p\, {dp \over dE}\, p^{2 - \sigma}\, E^b.
\end{equation}
The index $b$ will take on different values in the 
nonrelativistic (NR) and extreme-relativistic (ER) regimes, because
of the different dependence of $p(E)$.  
We approximate the distribution as two straight power-laws 
with energy indices $q$ and $s$, respectively, 
joined
at $m_p c^2$.  For the NR regime, $dp/dE = \sqrt{m/2E}$ and
\begin{equation}
K_i({\rm NR}) = 4 \pi K_p\, m^{(3 - \sigma)/2} \,2^{(1 - \sigma)/2}\, E^{s + (1 - \sigma)/2}
\ \Rightarrow \ q = (\sigma - 1)/2
\end{equation}
in order that $K_i$ be constant.  For the ER regime, $dp/dE = 1/c$ and
\begin{equation}
K_i({\rm ER}) = 4 \pi K_p \,c^{\sigma - 3}\,E^{s + 2 - \sigma}
\end{equation}
giving $s = \sigma - 2$ and $q = (s + 1)/2$.  Eliminating $K_p$, 
\begin{equation}
K_i({\rm NR}) = 2^{(1 - \sigma)/2} \left( m c^2 \right)^{(3 - \sigma)/2} K_i({\rm ER}).
\end{equation}
The total number density of accelerated protons is
\begin{equation}
N_i = \int_{E_l}^{m_pc^2} K_i({\rm NR}) E^{-q} dE
\end{equation}
where we have assumed that $s > 2$ so that we may neglect the relativistic
protons, and count as ``accelerated'' all those protons with energies
above $E_l$.  If the upstream number density is $n_H$, 
the number flux crossing the shock is $n_H u_{\rm shock}$, and
we assume a fraction $\zeta$ of those become accelerated protons, convecting
away at the postshock speed $u_2 = u_{\rm shock}/r$.  
This gives
\begin{equation}
\zeta \,n_H \,u_{\rm shock} = N_i u_2 
  = {u_2 \over q - 1} \,E_l^{1 - q}\, K_i ({\rm NR}).
\end{equation}
Finally, we obtain the coefficient of the assumed power-law distribution
of relativistic proton energies:
\begin{equation}
K_i({\rm ER}) = 2^{(s + 1)/2} (m_p c^2)^{(s - 1)/2}\,{s - 1 \over 2}\, r\, \zeta\,
  n_H\, E_l^{(s - 1)/2}.
\end{equation}
This formulation is slightly approximate; at $E = m_p c^2$, $N_i({\rm
  NR})/N_i({ER}) = 2^{-(s + 1)/2}$.  For $s = 2$, this factor is 2.83.
We take $E_l$ to be ten times the thermal momentum of protons, $E_l =
10 \,m_p \,u_{\rm shock}$.  The integrated fluxes we calculate just scale
with these uncertain parameters.
%but we simply absorb it into the definition of $\zeta$.

In R98, we considered three possible limitations to the maximum energy
particles could reach: finite remnant age, an abrupt increase in the
diffusion coefficient at some energy scale (basically a free
parameter), and, for electrons only, radiative losses.  Here we
neglect possible sudden changes in diffusive properties of the medium.
For the remnant ages we consider here, ions are essentially always
age-limited.  As in R98, we write the particle diffusion coefficient
as $\kappa_0$ for a parallel shock (${\bf B}$ parallel to the shock
normal) and $\kappa(\theta_{\rm Bn}) = \kappa_0 R_J(\theta_{\rm Bn})$,
with $\theta_{\rm Bn}$ the angle between the upstream shock normal and
{\bf B}, for arbitrary obliquity.  In the standard DSA assumptions of
Bohm-like diffusion, $\kappa_0 = \eta E c/3 e B$, with $\eta$ the
``gyrofactor'', the mean free path in units of the particle
gyroradius.  In the absence of turbulent rearrangement of {\bf B}, the
parallel component is unchanged while the tangential one is compressed
by $r$, so the upstream and downstream values of $B$ are related by
$B_2/B_1 \equiv r_B$ (with $1 \le r_B \le r$).  Furthermore, {\bf B}
is ``refracted'' so that $B_1 \cos\theta_{\rm Bn} = B_2 \cos\theta_{\rm
  Bn2}$, where $\theta_{\rm Bn2}$ is the downstream angle of {\bf B}
with respect to the shock normal.  Then $R_J$ is given in R98 as
\begin{equation}
  R_J \equiv {{\cos^2\theta_{\rm Bn} + \left[\sin^2\theta_{\rm Bn}/(1 + \eta^2)\right] + r(\cos\theta_{\rm Bn2}/\cos\theta_{\rm Bn}) \left\lbrace \cos^2 \theta_{\rm Bn2} + 
\left[ \sin^2 \theta_{\rm Bn2}/(1 + \eta^2)\right]\right\rbrace}\over {1 + r}}
\end{equation} 
% (correcting a typographical error in Equation 9 of R98).  For large
$\eta,$ $R_J$ can be much less than 1 \citep[much more rapid
  acceleration in perpendicular shocks;][]{jokipii87}, although both
obliquities produce slower acceleration than for smaller $\eta$.  If
turbulence is strong ($\eta \sim 1$), the distinction between parallel
and perpendicular shocks disappears.

Then R98 shows that the energy gain $dE$ in DSA in a time
$dt$ is given by
\begin{equation}
dE = {r - 1 \over r(r + 1)} \, {eB_1 \over \eta c} \, 
   {1 \over R_J} \, u_{\rm shock}^2 \,dt
\end{equation}
where $B_1$ is the upstream magnetic field.
We integrate this equation numerically in parallel with the
hydrodynamics to obtain $E_m({\rm age})$.  If the Sedov phase begins
at time $t_1$, the shock velocity $u_{\rm shock}(t) = u_{\rm shock}(t_1) (t/t_1)^{-0.6}$. Then
the time-dependence of $E_m({\rm age})$ is given by
\begin{equation}
\int_{t_1}^t u_{\rm shock}^2\, dt 
   = 5 [u_{\rm shock}(t_1)]^2\, t_1 \left[1 - \left(t \over t_1\right)^{-0.2}\right]
\end{equation}
which is nearly constant. 

In the evolutionary stages we consider, electron energies are always
limited by losses.  If $B > 3.3 \ \mu$G, as is always the case here
(except for the region near the poles in model SA; see below),
losses are dominated by synchrotron (rather than inverse-Compton
upscattering of cosmic microwave background photons, ICCMB).  R98
gives the maximum energy as
\begin{equation}
E_m({\rm loss}) = 0.32 \, C_l(r, \theta_{Bn})(\eta R_J B_1)^{-1/2} 
   {u_{\rm shock} \over 10^8 \ {\rm cm \ s}^{-1}}
\end{equation}
where $C_l(r, \theta_{\rm Bn})$ is a factor of order unity depending on
the compression ratio and obliquity angle $\theta_{\rm Bn}$.  

We then assume our particle distributions immediately behind the
shock are given by
\begin{equation}
N_i = K_i({\rm ER})\, E^{-s}\, e^{-E/E_{mi}} \ \ ({\rm ions}) \ \ {\rm and} \ \ 
N_e = k_{\rm ei}\, K_i({\rm ER})\, E^{-s}\, e^{-E/E_{me}} \ \ ({\rm electrons})
\end{equation}
with $k_{ei} = 0.02$, and where $E_{\rm mi}$ and $E_{\rm me}$ are the
lowest maximum energy applicable to ions or electrons, respectively -- 
but as mentioned above, in the current work $E_{\rm mi}$ is always
due to the age limitation and $E_{\rm me}$ to losses.

\subsection{Magnetic-field evolution}

In our limit, the magnetic and cosmic-ray energy densities are small
compared to the post-shock pressure.  We consider two cases for the
magnetic field: a constant magnetic field (uniform in strength and
direction) for the analytic Sedov case, and a magnetic field assumed
to be isotropized with a strength proportional to the square root of
gas density (both upstream and downstream) in the numerical cases.  No
magnetic-field amplification is assumed.

In the analytic-dynamics case, it is possible to follow the evolution
of the radial and tangential components of magnetic field separately,
under the assumption of flux freezing (no further turbulent or
cosmic-ray amplification).  If the immediate postshock radial and
tangential components of {\bf B} are $B_{2r}$ and $B_{2t}$, and
$r_i(r)$ is the radius at which a fluid element currently at $r$ was
shocked, then $B_r = B_{2r}(r_i/r)^2$ and $B_t =
B_{2t}(\rho(r)/\rho_2) (r/r_i)$, with $\rho_2$ the immediate postshock
density \citep{duin75,rc81}.  This means that while the dynamics in
this case is one-dimensional, the images reflect the anisotropic
evolution of magnetic field and will depend on the aspect angle $\phi$
between the upstream magnetic field, assumed constant in this case,
and the line of sight.  In the numerical cases, the magnetic field is
assumed isotropized everywhere, so the predicted images are also
spherically symmetric.  Again, the analytic case is similar to that
described in R98.

\section{Radiative processes}

Synchrotron emission is calculated from standard formulae
\citep[e.g.,][]{pacholczyk70}, again as in R98.  To calculate the
contributions to the intensity by inverse-Compton scattering,
nonthermal electron-ion bremsstrahlung, nonthermal electron-electron
bremsstrahlung, and neutral pion decay, we implemented the
emissivities outlined in \citet{houck06}, with one difference. Instead
of the exact (to first order in perturbation theory) differential
cross-section for electron-electron of \citet{haug75}, we used the
approximate parameterization of \citet{baring99}.  We made this choice
because the expression for the electron-electron cross-section of
\citet{haug75} is awkward and computationally expensive. On the other
hand, the approximation of \cite{baring99} is straightforward to
implement, requires less computational effort, and agrees with the
exact result to better than 10 \% for electron energies $\gtrsim 5$
MeV (or equivalently, $\gamma_e \gtrsim 10$) \citep{baring99}. Since
the electrons responsible for gamma-ray emission in our simulations
have $E_e \gg 5$ MeV, this prescription is more than suitable for our
purposes.

\begin{figure}
\includegraphics[scale=0.8]{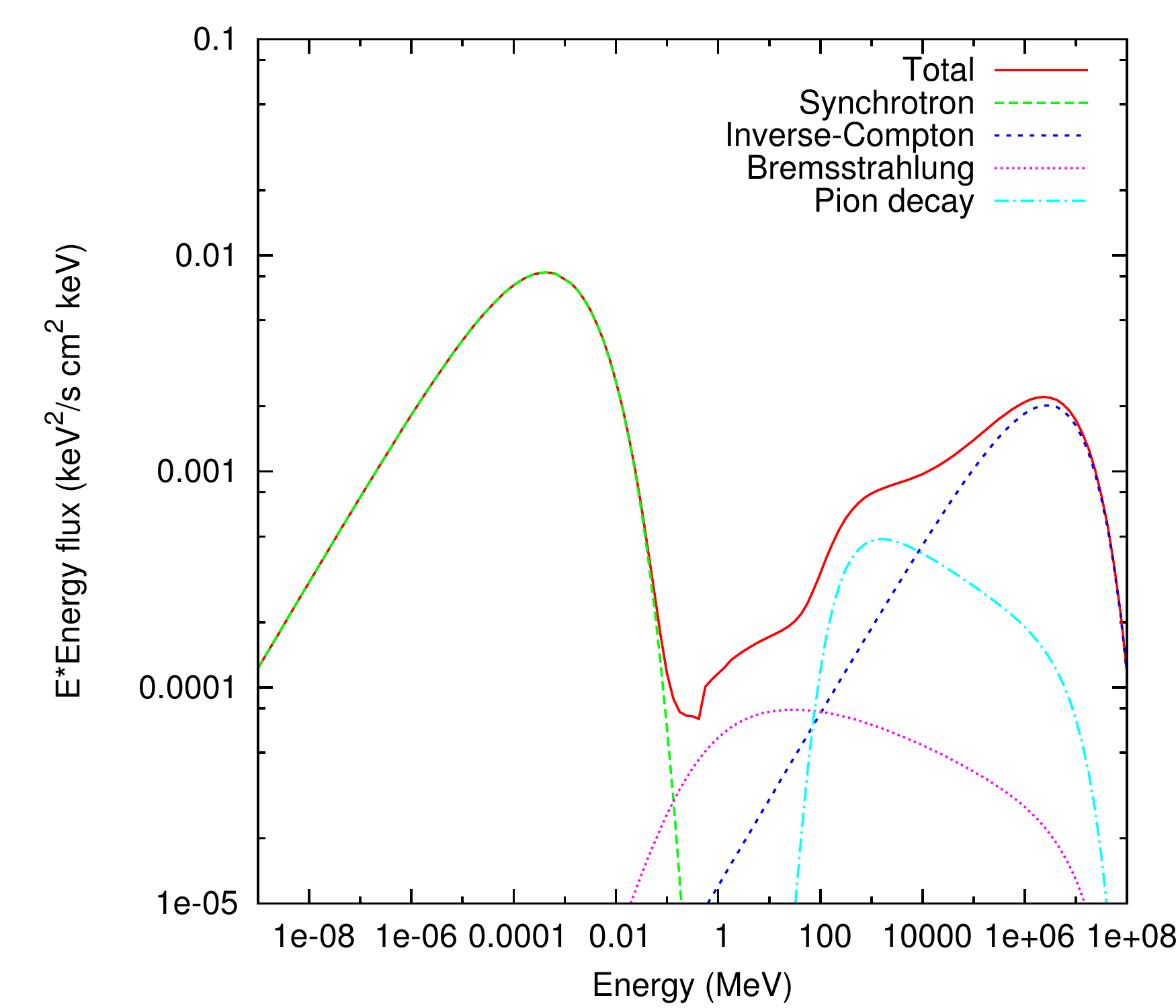}
\caption{Emissivities for the four principal processes producing X-ray
  to gamma-ray emission.  Parameters are shown in
  Table~\ref{uniparameters}.  }
\label{homogen}
\end{figure}

Figure~\ref{homogen} shows the emissivities for the four processes
(given as fluxes from a homogeneous source).  (Electron-electron
bremsstrahlung, negligible for nonrelativistic electrons, is included
but not shown; it tracks electron-ion bremsstrahlung but weaker by a
factor of a few.)  The parameters, shown in Table~\ref{uniparameters},
are chosen to match those for a Sedov model described below.  In a
one-zone (``zero-D'') model, the fluxes simply scale as these
emissivities.  Fig.~\ref{homogen} shows fluxes assuming a spherical
remnant with a specified radius, and a volume filling factor for
emission of 1/4, at a distance of 5 kpc.  The radius chosen for the
fluxes in Fig.~\ref{homogen} is $2.5 \times 10^{19}$ cm for comparison
with other models (see below).

\begin{deluxetable}{lccccccccc}
\tablecolumns{10}
\tablewidth{0pc}
%\tabletypesize{\scriptsize}
\tablecaption{Model Parameters: Uniform Medium}

\tablehead{
\colhead{Model} & $n_0$ & $s$ & $B_2$ & $\zeta$  & $R_s$ & $t$ & $u_s$ & $E_{me}$&
$E_{mi}$ \\
\colhead{}   & (cm$^{-3}$) & & ($\mu$G) & & (cm) & (yr) & (km s$^{-1}$) & (TeV) & (TeV)}
\startdata
%\\

Homogeneous & 0.97\tablenotemark{a} & 2.2 & 4.5 & $10^{-4}$ 
 &$2.5 \times 10^{19}$ &1700 &1850 & 54 & 137 \\

Sedov analytic (SA)& 0.25 & 2.2 & 4.5\tablenotemark{b} &  $10^{-4}$ &$2.5 \times 10^{19}$ & 1700
 &1850 & 51 & 429 \\

U1 & 0.25  & 2.2 & 4.5 & $10^{-4}$ 
 &$2.0 \times 10^{19}$ &972 &2590 & 76 & 123 \\

U2\tablenotemark{c} & 0.25 & 2.2 & 4.5  & $10^{-4}$ &$2.5 \times 10^{19}$ & 1700
 &1850 & 54 & 137 \\

U3 & 0.25  & 2.2 & 4.5 & $10^{-4}$ 
 &$6.6 \times 10^{19}$ &21,000 &427 & 43 & 203 \\

\enddata \tablecomments{Models U1, U2, and U3 use the 1-D hydrodynamic simulations.}
\tablenotetext{a}{Corresponds to downstream density in all other models.}
\tablenotetext{b}{Average over obliquities.  Upstream field $B_1$ is $1.4\ \mu$G.}
\tablenotetext{c}{Same as model JB in standard-collision set.}
\label{uniparameters}
\end{deluxetable}

\begin{deluxetable}{lccccccccc}
\tablecolumns{10}
\tablewidth{0pc}
%\tabletypesize{\scriptsize}
\tablecaption{Model Parameters: Standard Collision}

\tablehead{
\colhead{Model} & $n_0$ & $s$ & $B_2$ & $\zeta$  & $R_s$ & $t$ & $u_s$ & $E_{me}$&
$E_{mi}$ \\
\colhead{}   & (cm$^{-3}$) & & ($\mu$G) & & (cm) & (yr) & (km s$^{-1}$) & (TeV) & (TeV)}
\startdata
%\\

Justbefore (JB) & 0.25\tablenotemark{a} & 2.2 & 4.5  & $10^{-4}$ &$2.50 \times 10^{19}$ & 1700
 &1850 & 54 & 137 \\
A1  & 5.0\tablenotemark{b} & 2.2 & 20  & $10^{-4}$ &$2.67 \times 10^{19}$ 
 &2027 &737  & 22 & 141 \\
A2  & 5.0\tablenotemark{b} & 2.2 & 20  & $10^{-4}$ &$2.71 \times 10^{19}$ 
 &2236 &586  & 14 & 141 \\
A3  & 5.0\tablenotemark{b} & 2.2 & 20  & $10^{-4}$ &$2.80 \times 10^{19}$ 
 &2740 &499  & 12 & 142 \\
A4  & 5.0\tablenotemark{b} & 2.2 & 20  & $10^{-4}$ &$2.92 \times 10^{19}$ 
 &3541 &413  & 9.7 & 143 \\
A5  & 5.0\tablenotemark{b} & 2.2 & 20  & $10^{-4}$ &$3.13 \times 10^{19}$ 
 &5350 &318  & 7.4 & 143 \\

\enddata \tablecomments{In all cavity models, the wall is located at
  $R_w = 2.67 \times 10^{19}$ cm, with a density jump there of a
  factor of 20, unless otherwise noted.  The collision occurs at a time
of about 2000 yr, when the shock velocity is 1690 km s$^{-1}$.}
\tablenotetext{a}{Density in cavity interior.}
\tablenotetext{b}{Density in cavity wall.}
\label{scparameters}
\end{deluxetable}

\section{Results}

Images and spectra were calculated for both classes of model: those
encountering a uniform medium and those interacting with a cavity
wall.  In the former case, the remnant dynamics are just those of a
Sedov blast wave.  Two types of Sedov models are described: one using
the analytical Sedov solution and including anisotropies due to
magnetic-field variations; and one suitable for the hydrodynamic
simulations, in which the magnetic field is assumed isotropized with a
strength simply proportional to the square root of density at all
times.  The former model will be referred to as Sedov, and the latter,
models U1, U2, and U3 of Table~\ref{uniparameters}.  (Model U2 will
be described in particular; it has an age corresponding to
a time just before collision with the wall, and is also listed as JB (``just
before'') in Table~\ref{scparameters}.)

For cavity models, we focus on a ``standard collision'' (SC) scenario,
with model parameters given in Table~\ref{scparameters}.  We exhibit
results at five times after the collision.  More extensive parameter
studies can be found in Tang (PhD thesis, in preparation).

\subsection{Dynamics}

\begin{figure}
\centerline{\includegraphics[scale=0.5]{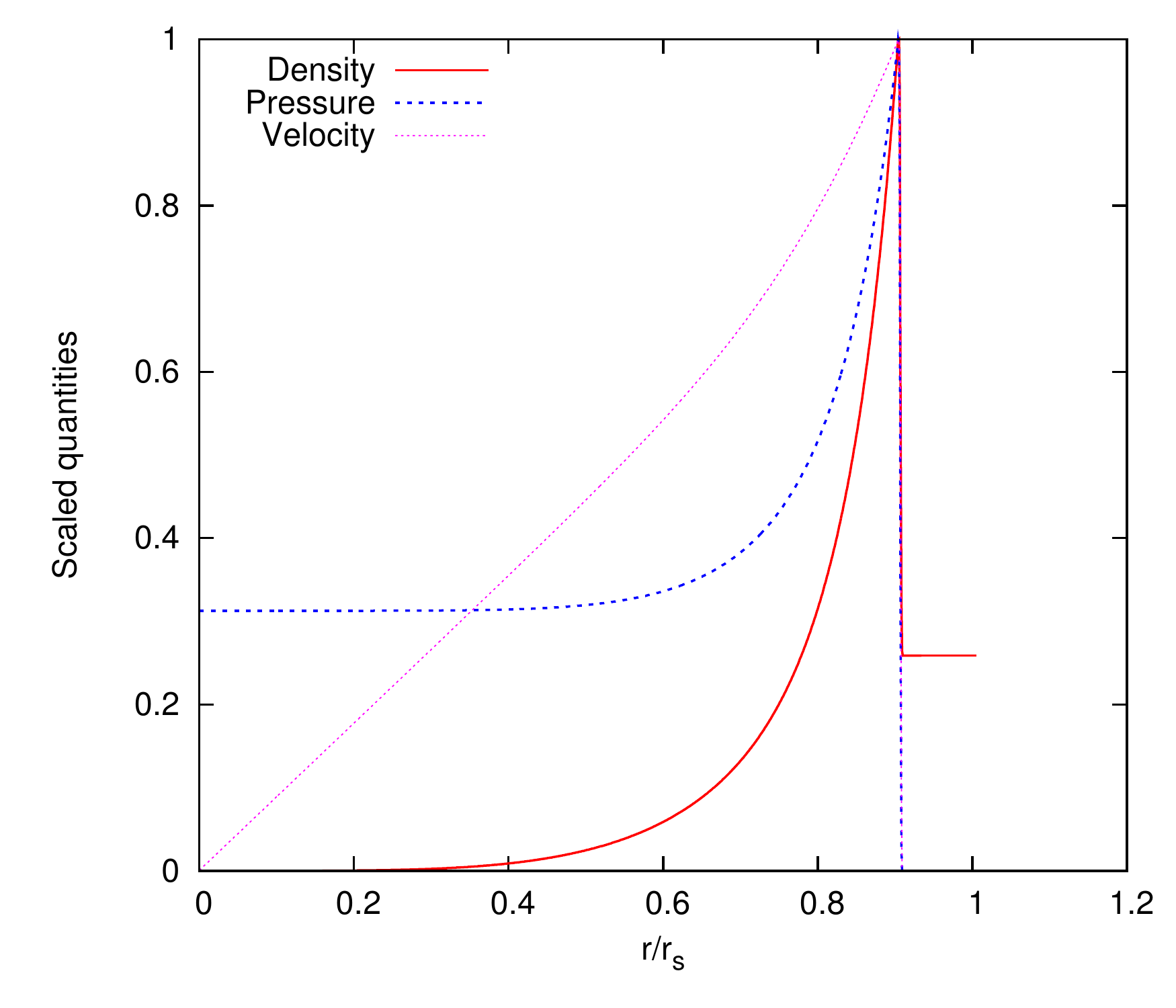}\hskip0.1truein
   \includegraphics[scale=0.5]{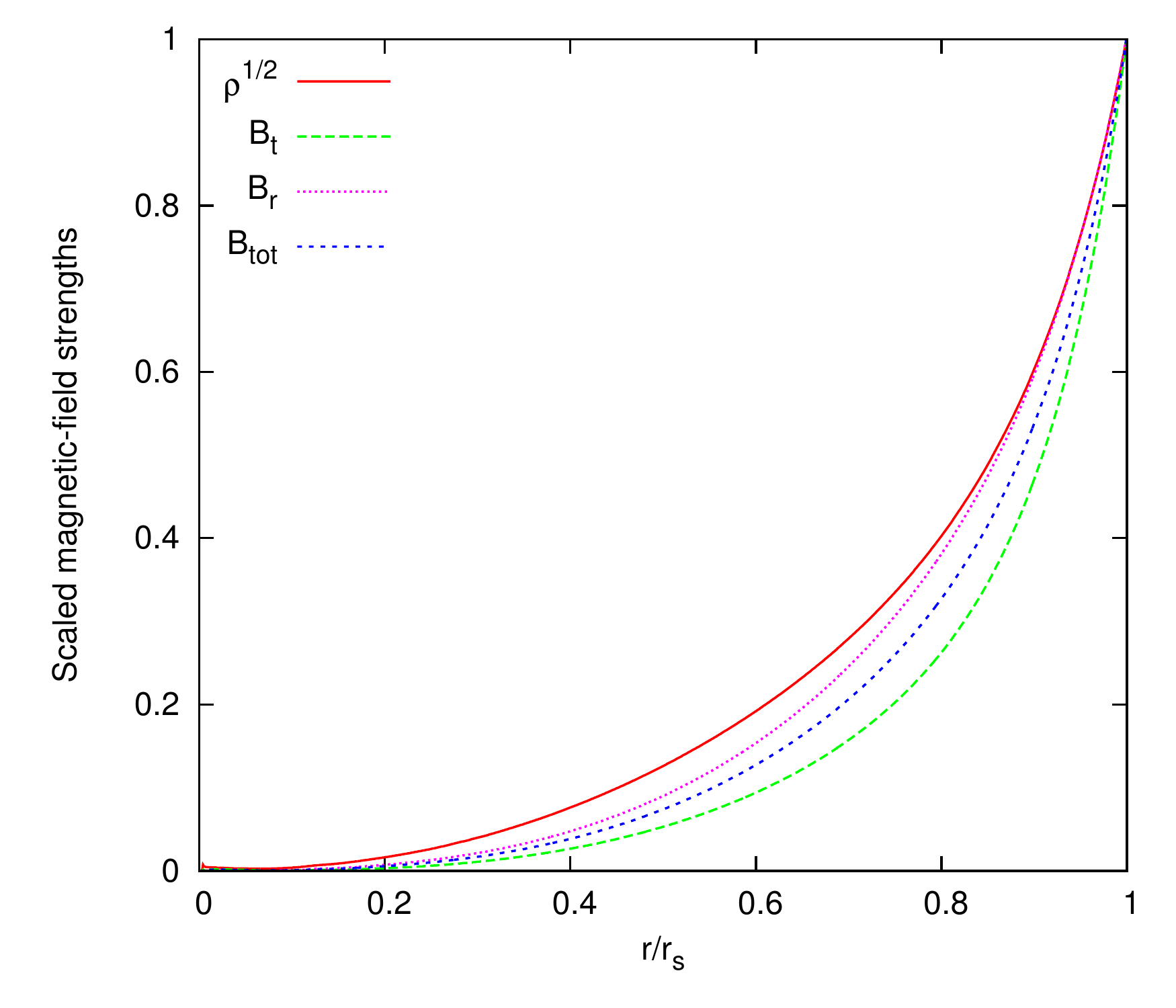}}
\caption{Left: Scaled profiles of dynamical quantities for Sedov
  dynamics.  Right: Scaled profiles of magnetic-field components for
  Sedov density profiles.  The total magnetic-field strength assuming
  evolution only due to flux-freezing is the blue (short-dashed) curve,
  while the approximation of $B \propto \rho^{1/2}$ is shown in red
  (top, solid curve).  Note that the latter is always larger than the
  former.}
\label{Sedov}
\end{figure}

The familiar Sedov dynamical quantities and the radial dependence of
magnetic fields are plotted in Figure~\ref{Sedov}.  The field strength
governs the electron distribution (through synchrotron losses), and
therefore affects the inverse-Compton emissivity as well as the
synchrotron emissivity.  The total strength assuming flux-freezing
(dashed, or blue, or third curve from the top in Fig.~\ref{Sedov}) is
always smaller than that from the simple assumption $B^2 \propto
\rho$, used in the hydrodynamic simulations.

\subsection{Emission from Sedov-phase remnants}

We first compare the analytic Sedov model (SA) with the homogeneous
(zero-D) model, then with the numerical model U2.  

\subsubsection{Spectra}

\begin{figure}
\centerline{\includegraphics[scale=0.5]{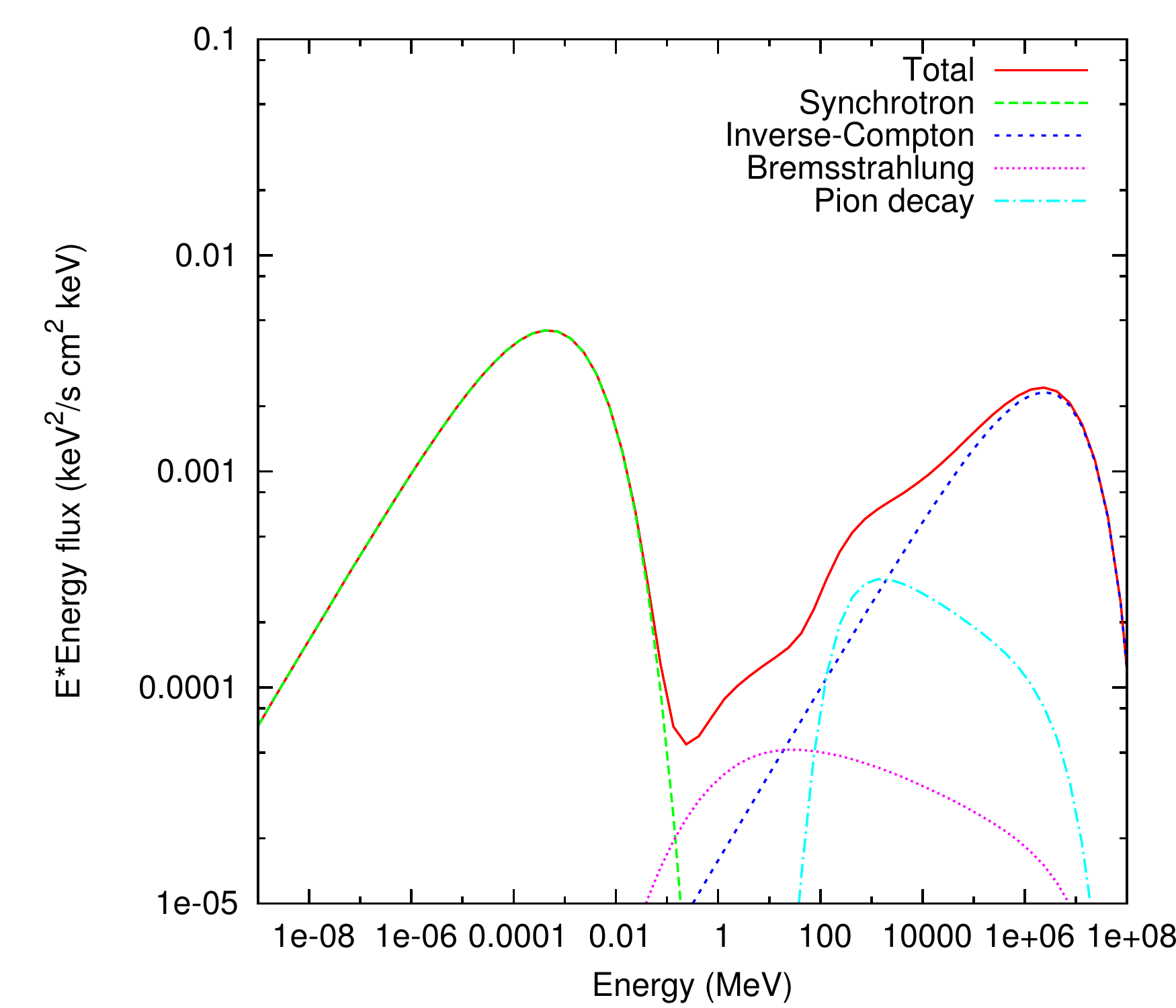} \hskip0.1truein
   \includegraphics[scale=0.5]{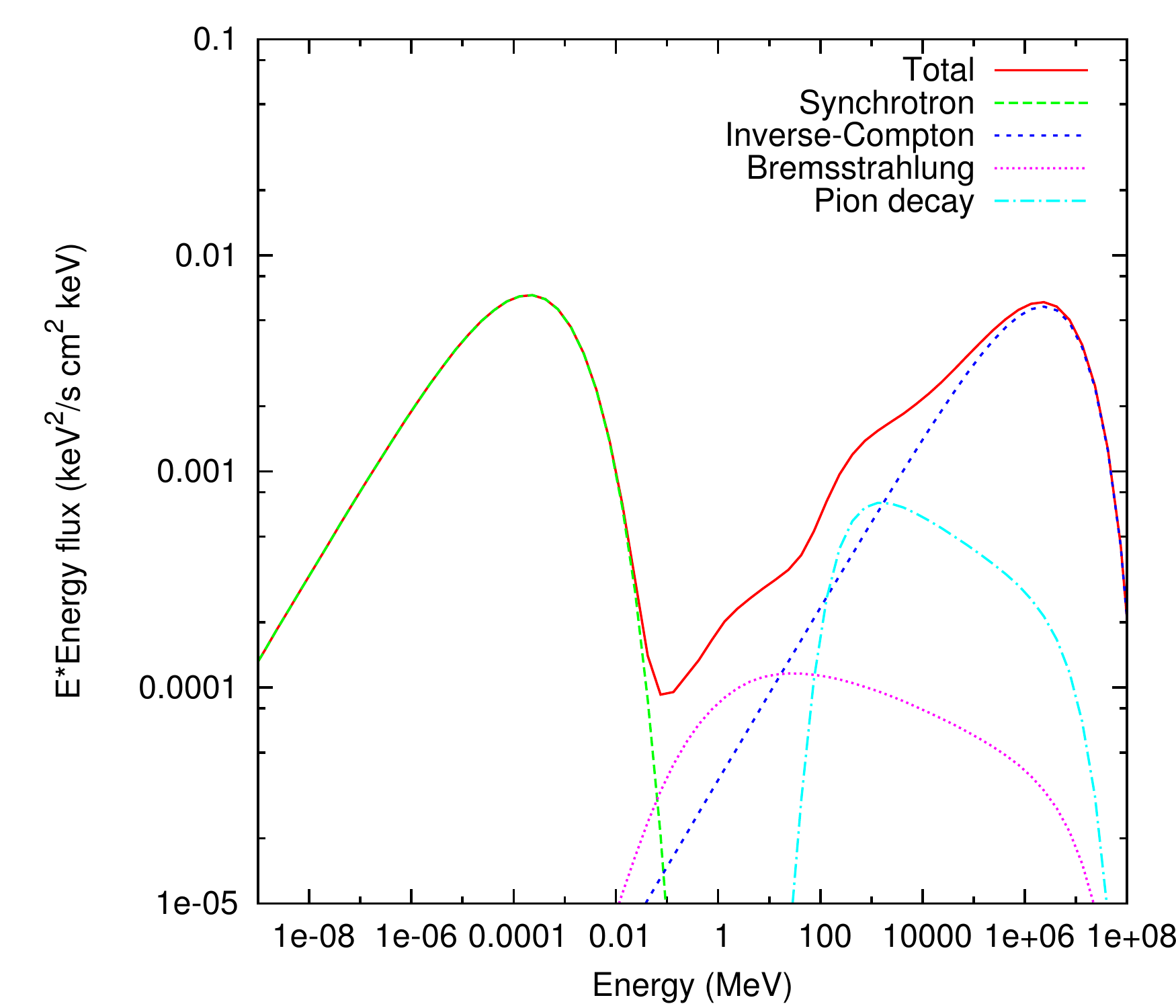}}
\caption{Left: Fluxes for a model based on Sedov analytic dynamics (parameters
shown in Table~\ref{uniparameters}).  The magnetic-field spatial variation
is the primary reason for differences between this model and the
one-zone (zero-D) spectra (Fig.~\ref{homogen}). See text for details.
Right: Fluxes for the numerical Sedov model with the same input parameters,
differing only in the treatment of magnetic field.
}
\label{sedovannum}
\end{figure}

Figure~\ref{homogen} shows spectral fluxes for a
homogeneous, spherical remnant with properties shown in
Table~\ref{scparameters}, to be compared with Figure~\ref{sedovannum}.
The evolution of magnetic-field strength in the latter model brings
about differences of factors of $\sim 3$: while the synchrotron flux
at low frequencies is very close to that from a one-zone model, the
cutoff frequency (actually a function of position in the SA model) is
higher by about a factor of 3.  All the gamma-ray fluxes are lower in
the SA model by factors of about 3 (a slightly larger factor for
inverse-Compton), due to the density dropoff behind the shock and the
$n^2$ density dependences of those processes.  The
combination of higher rolloff frequency for the synchrotron component
and lower IC normalization makes the ratio of peak synchrotron and IC
fluxes differ considerably from that in the one-zone model.

The hydrodynamic model before the blast wave encounters the cavity
wall (model U2, or JB) is also basically a Sedov model, but differs in the
treatment of magnetic field.  As Fig.~\ref{Sedov} shows, that
difference in the remnant interior is substantial.
Figure~\ref{sedovannum} compares the spatially integrated fluxes in
the two models.  The differences are surprisingly small; the
hydrodynamic model does have a higher synchrotron flux, as expected
from the somewhat higher magnetic field in the remnant interior.  But
all flux differences are of order a factor 3 or less.

\subsubsection{Images}

The anisotropy of the magnetic field in the SA model allows departure
from circular symmetry in the images, if not in the dynamics, since
the magnetic-field strength affects the maximum energies to which
electrons or ions can be accelerated.  See \cite{reynolds08} for a
review. For electrons, if synchrotron losses limit the maximum energy
(as is the case for the relatively mature cases considered here), the
maximum electron energy $E_{me}$ (actually the $e$-folding energy of a
roughly exponential cutoff to the power-law, in simple cases) obeys
$E_{me} \propto B^{-1/2}$, so the characteristic synchrotron frequency
emitted by such electrons, $\nu_c$, obeys $\nu_c \propto E_{me}^2 B $
which is independent of $B$.  However, the strength of the synchrotron
component does of course depend on the magnetic-field strength.  For
ions, the most likely mechanism limiting acceleration is the finite
size or age of the remnant (roughly equivalent criteria), resulting
in $E_{mi} \propto B$ -- so higher ion energies can be achieved where
the magnetic field is higher.  However, higher fields produce {\sl lower}
electron maximum energies since those are loss-limited for the models
considered here.  

\begin{figure}
\centerline{\includegraphics[scale=0.9]{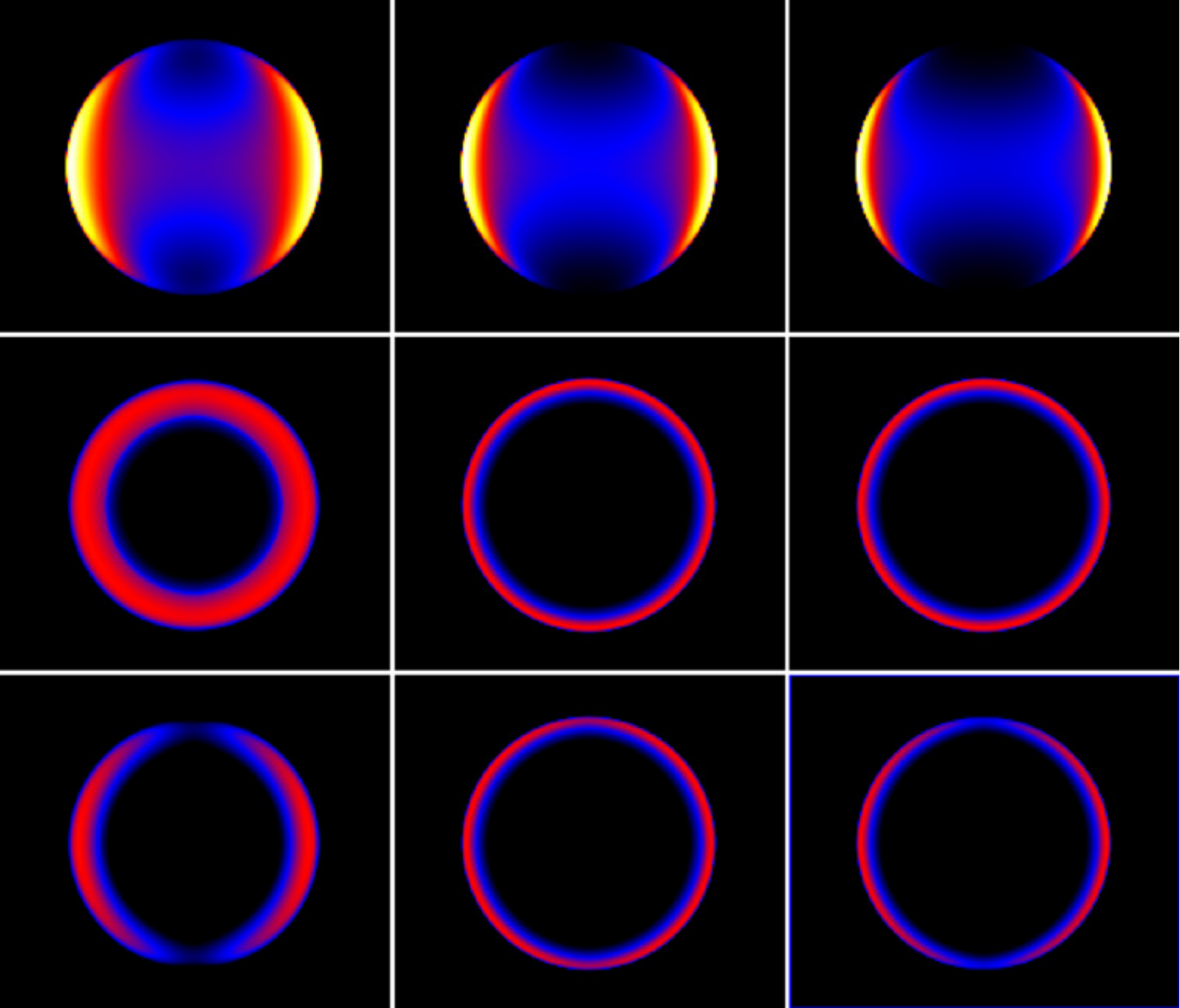}}
\caption{Images of SA model due to different radiative processes, at
  different photon energies.  Top row: Synchrotron emission at 1 GHz,
  1 keV, and 10 keV.  Note the thinning of the rim at higher energies.
  Middle row: IC, bremsstrahlung, and $\pi^0$ emission at 1 GeV.  The
  IC shell is broader because the photon target field is uniform,
  while both bremsstrahlung and $\pi^0$ emission targets also scale
  with density, so the emissivity varies as density squared in the
  interior.  (The apparent smaller size of the IC image is an
  illusion.)  Bottom row: Same processes at 10 TeV.  The magnetic
  field is lowest ($1.4\ \mu$G) at position angles 0 and $180^\circ$,
  so maximum ion energy is lowest there.  However, $B$ there is so low
  that the energy density in CMB photons dominates over magnetic-field
  energy density, and electron energy losses are dominated by ICCMB
  interactions.  This reduces the maximum electron energy, causing the
  lower IC emission there.}
\label{SA9ims}
\end{figure}

Images for the SA model are shown in Fig.~\ref{SA9ims}.  The uniform
upstream magnetic field is in the vertical direction in the plane of
the figures.  The synchrotron emission is brightest near the equators,
in a ``belt'' (as opposed to a ``cap'') geometry, simply due to higher
synchrotron emissivity where the magnetic field is larger due to
compression of the tangential component.  The synchrotron emission is
confined to a thin region, that becomes thinner for higher photon
energies due to downstream energy losses by the electrons (the ``thin
rims'' phenomenon; see R08).  The 1 GeV images of IC, electron-ion
bremsstrahlung, and $\pi^0$ emission are all circularly symmetric.
Since the IC emission represents interaction of the electrons with a
uniform photon sea (the CMB), there is a slower radial dropoff of
emission toward the center than is true for either of the other
processes, in which both primary and target populations drop steeply
in density toward the remnant center, resulting in a density-squared
dependence.  (The synchrotron emission varies as the electron density
times $B^{1+\alpha}$, and with $\alpha = 0.6$ and $B \propto
\rho^{1/2}$ in the numerical models, synchrotron emission is also
roughly proportional to density squared.)

\begin{figure}
\centerline{\includegraphics[scale=0.5]{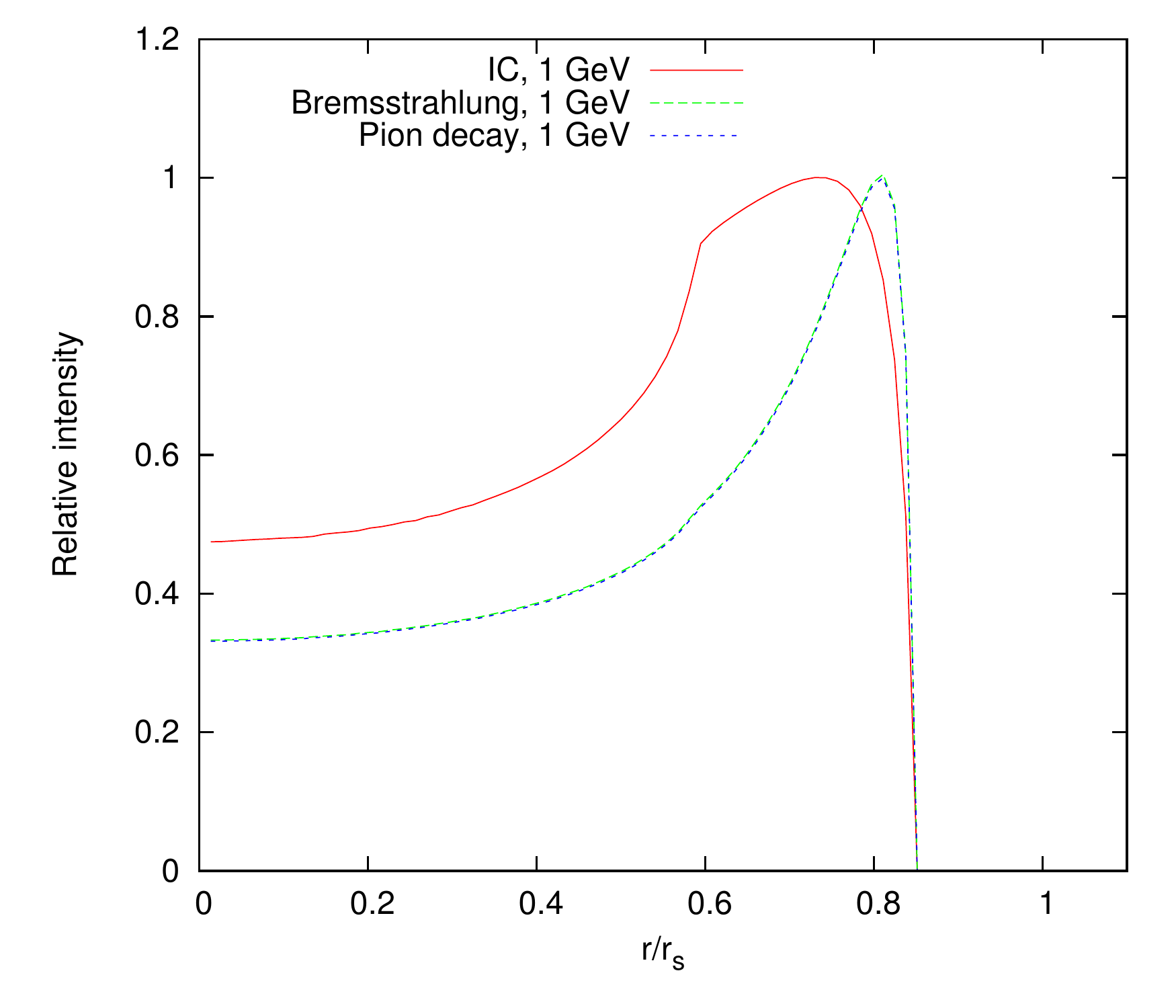}
   \hskip0.2truein \includegraphics[scale=0.5]{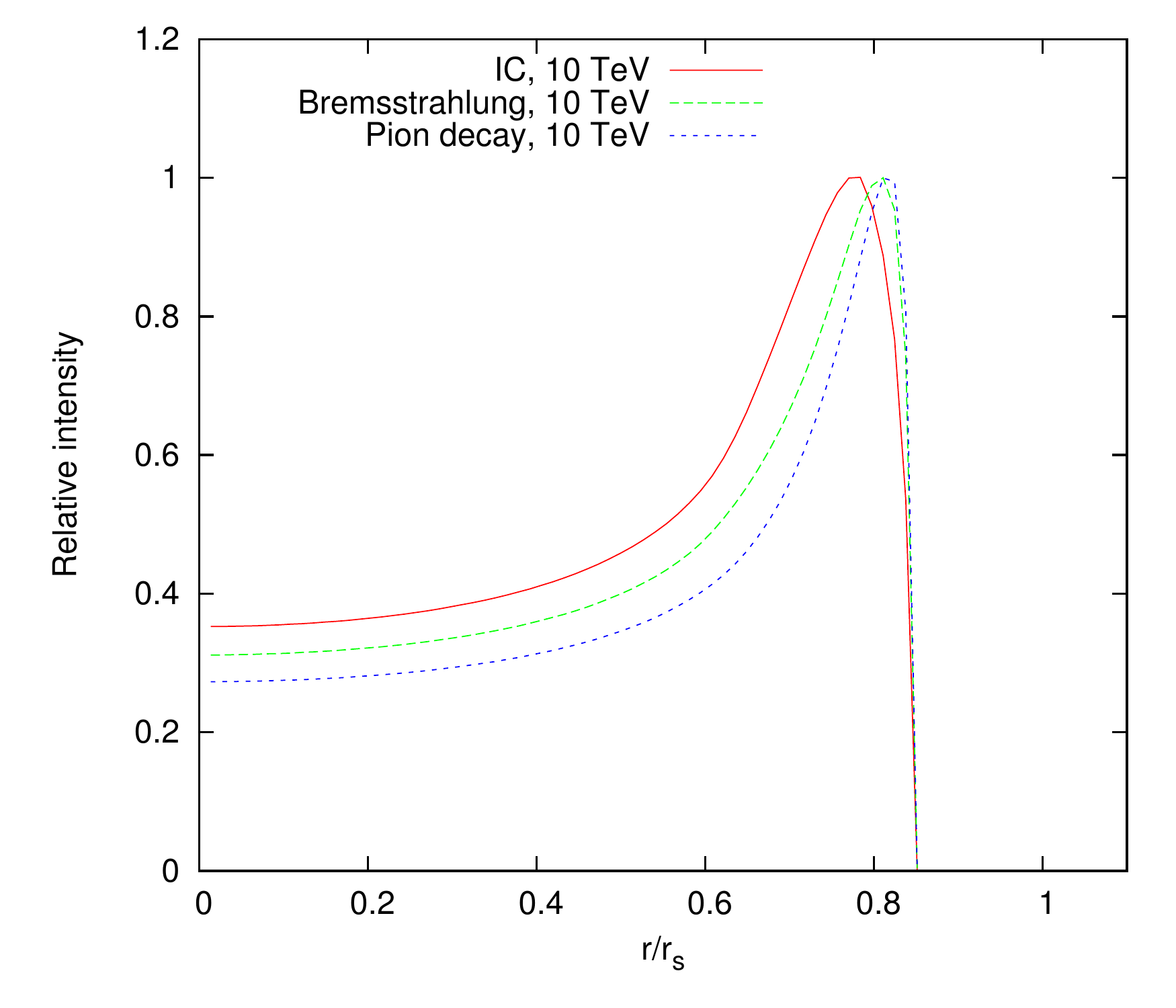}}
\caption{Left: Radial profiles (along a horizontal or equatorial
  radius) of the three 1 GeV images in Fig.~\ref{SA9ims}.  Right:
  Same, for 10 TeV images.  The bremsstrahlung and $\pi^0$ profiles
  are virtually identical at 1 GeV, but the broader IC profile might
  make it distinguishable observationally.}
\label{SAprofiles}
\end{figure}

At 10 TeV, the lower post-shock magnetic field at the poles results in
slightly lower maximum ion energies, reducing emission there.
Electron maximum energies would be higher (Equation (13)), but the
field drops below the value at which the magnetic energy density
equals that in CMB photons, so there is a ceiling to electron
energies, and the shell emission has a minimum at the poles for very
high photon energies.  That minimum disappears for larger
magnetic-field strengths.  Fig.~\ref{SAprofiles} shows radial profiles
of those images.  The inflection in the IC profile is an artifact, but
the general property of much broader profiles in IC than in pion-decay
emission is robust and true over a wide range of energies.

\subsection{Time evolution}

\begin{figure}
\centerline{\includegraphics[scale=0.4]{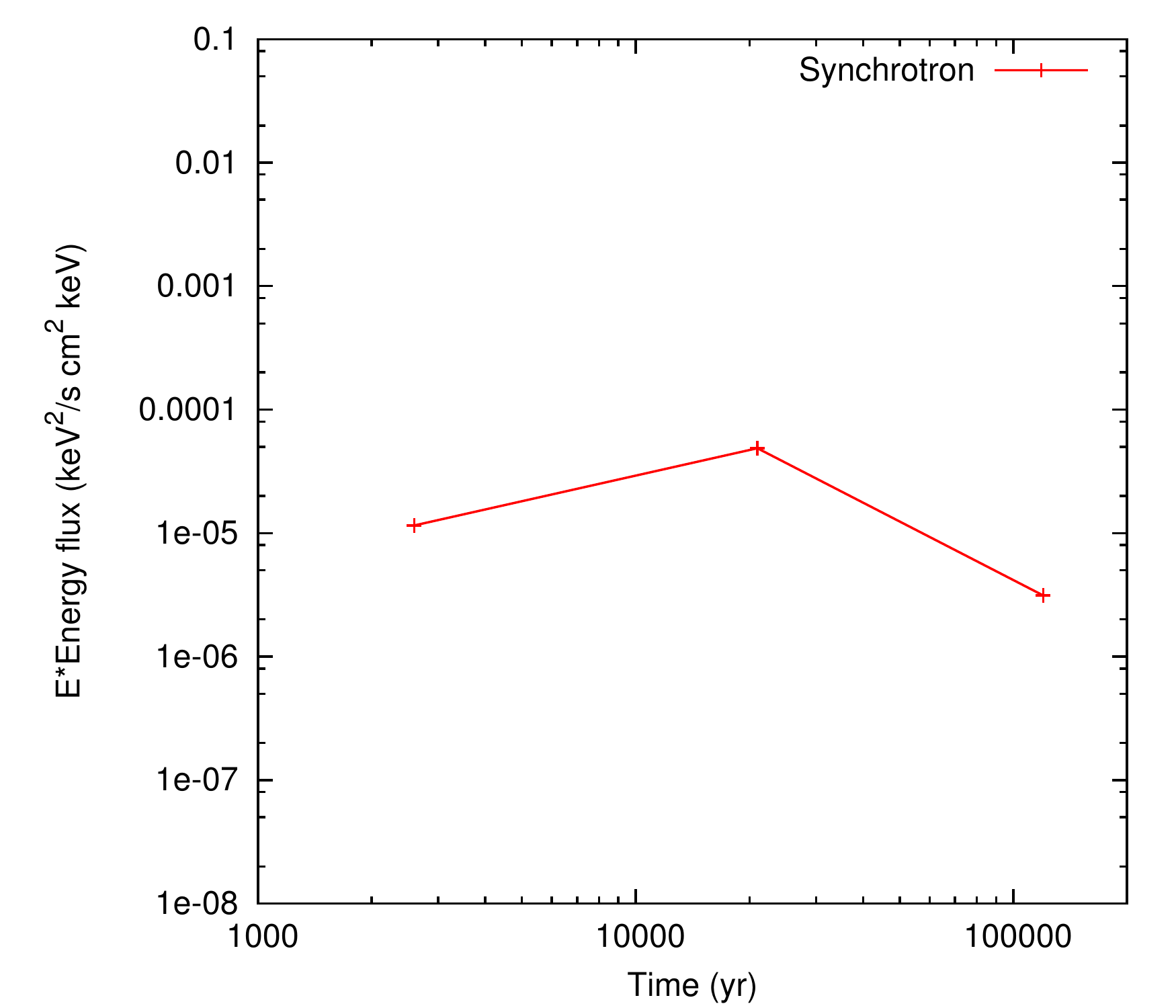}\hskip10truept
  \includegraphics[scale=0.4]{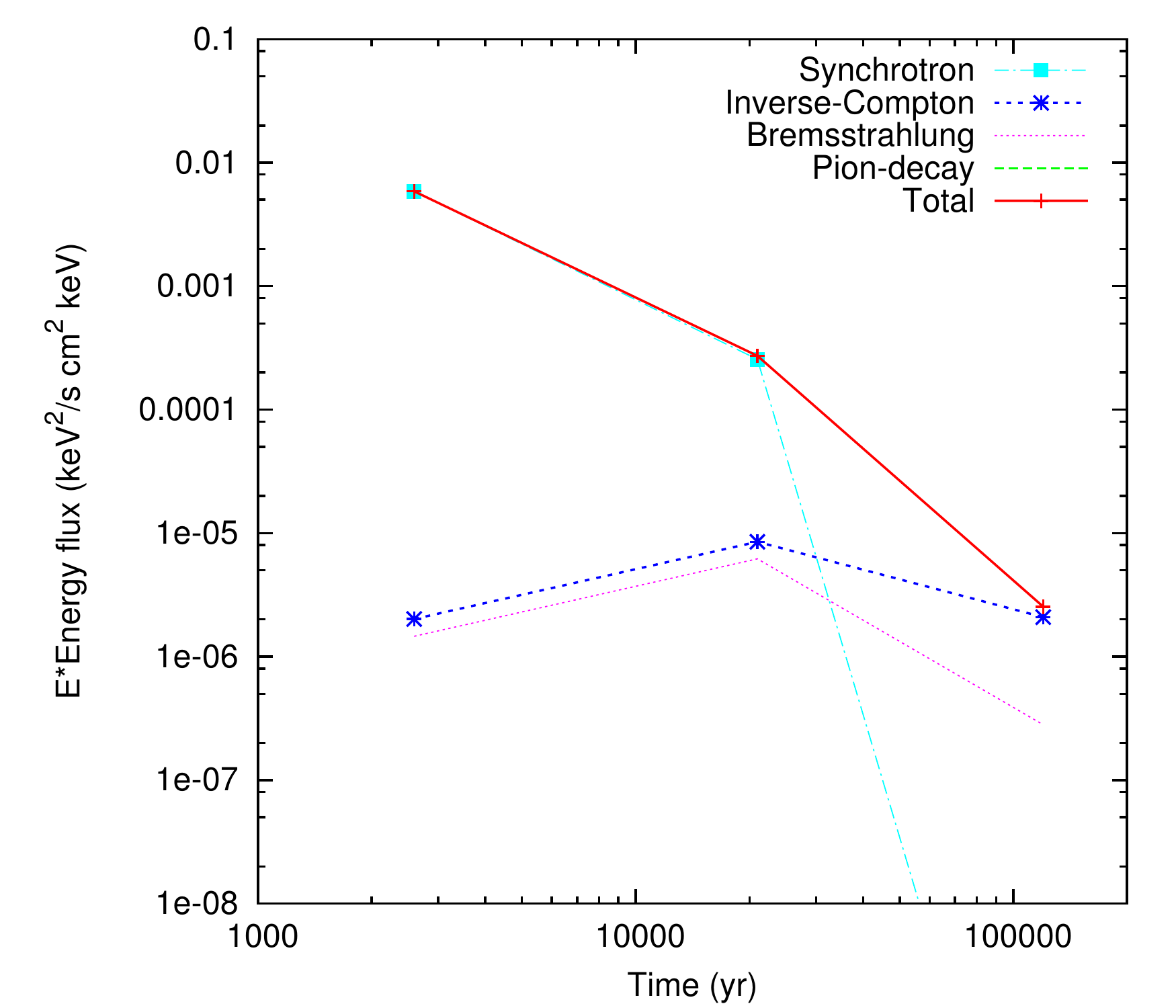}}
\centerline{\includegraphics[scale=0.4]{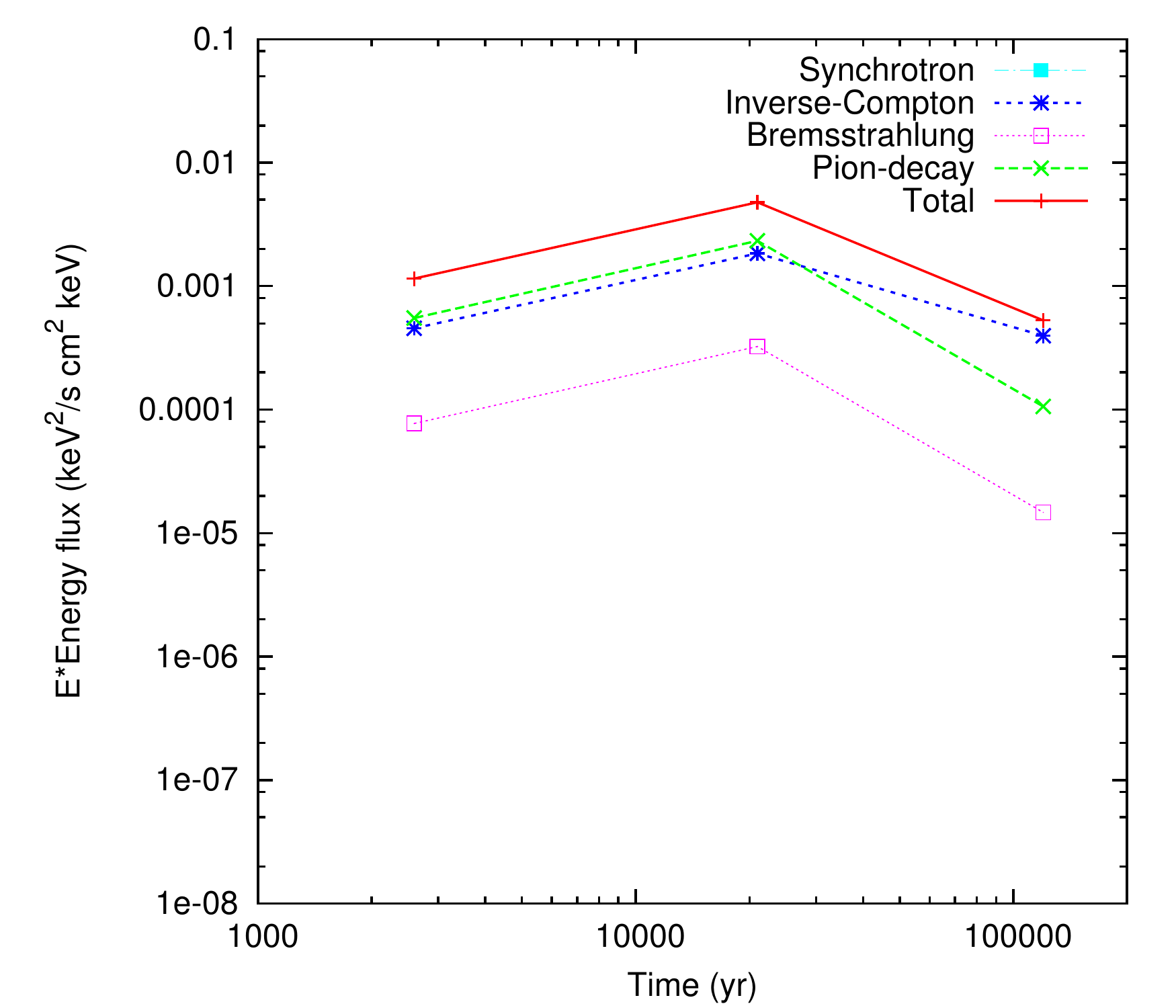}\hskip10truept
  \includegraphics[scale=0.4]{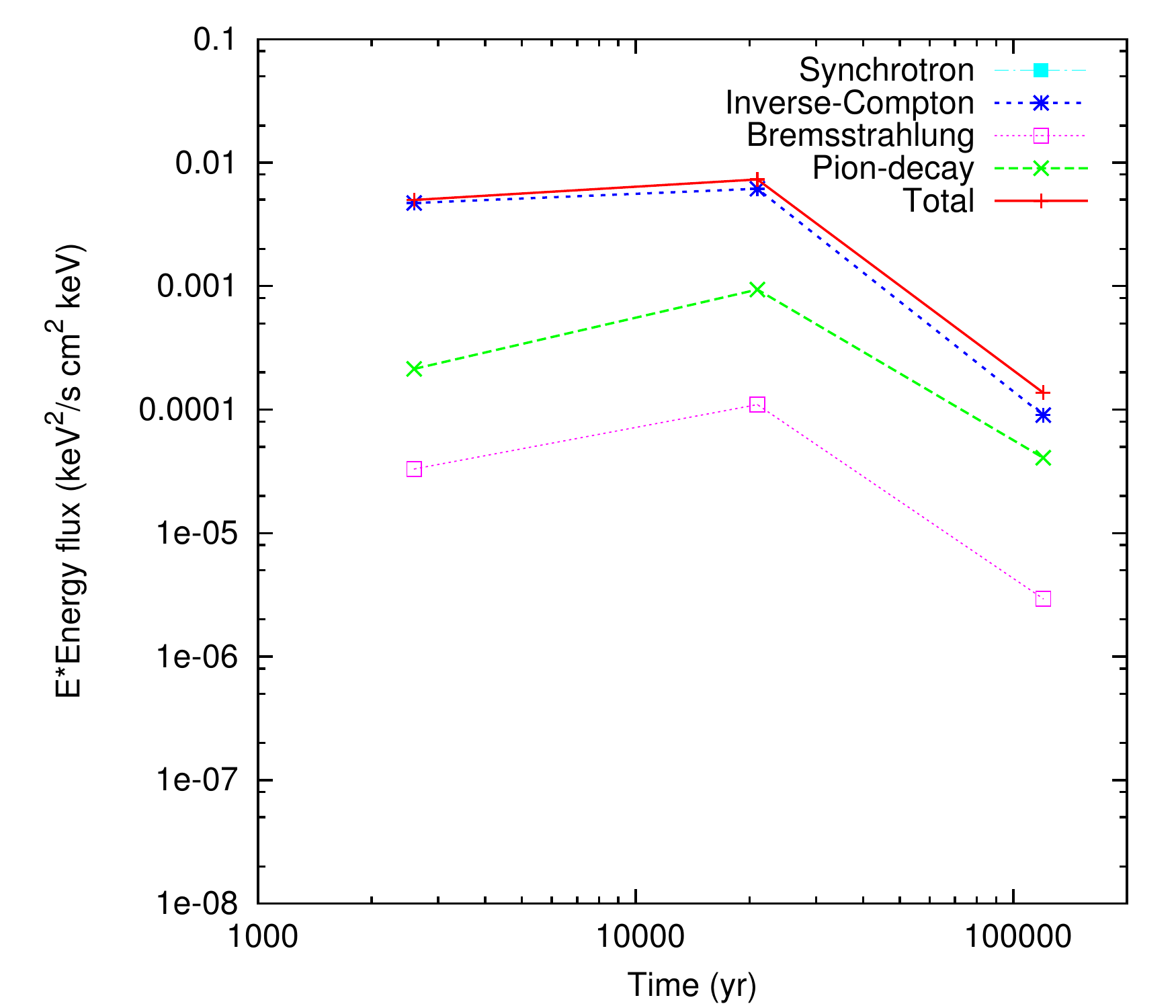}}
\caption{Fluxes $E^2 F(E)$ as a function of time, for the uniform-density model.
  Upper left: 1 GHz.  Upper right: 1.33 keV.  Lower left: 1 GeV.
  Lower right: 1 TeV.}
\label{unifluxes}
\end{figure}

\begin{figure}
\centerline{\includegraphics[scale=0.8]{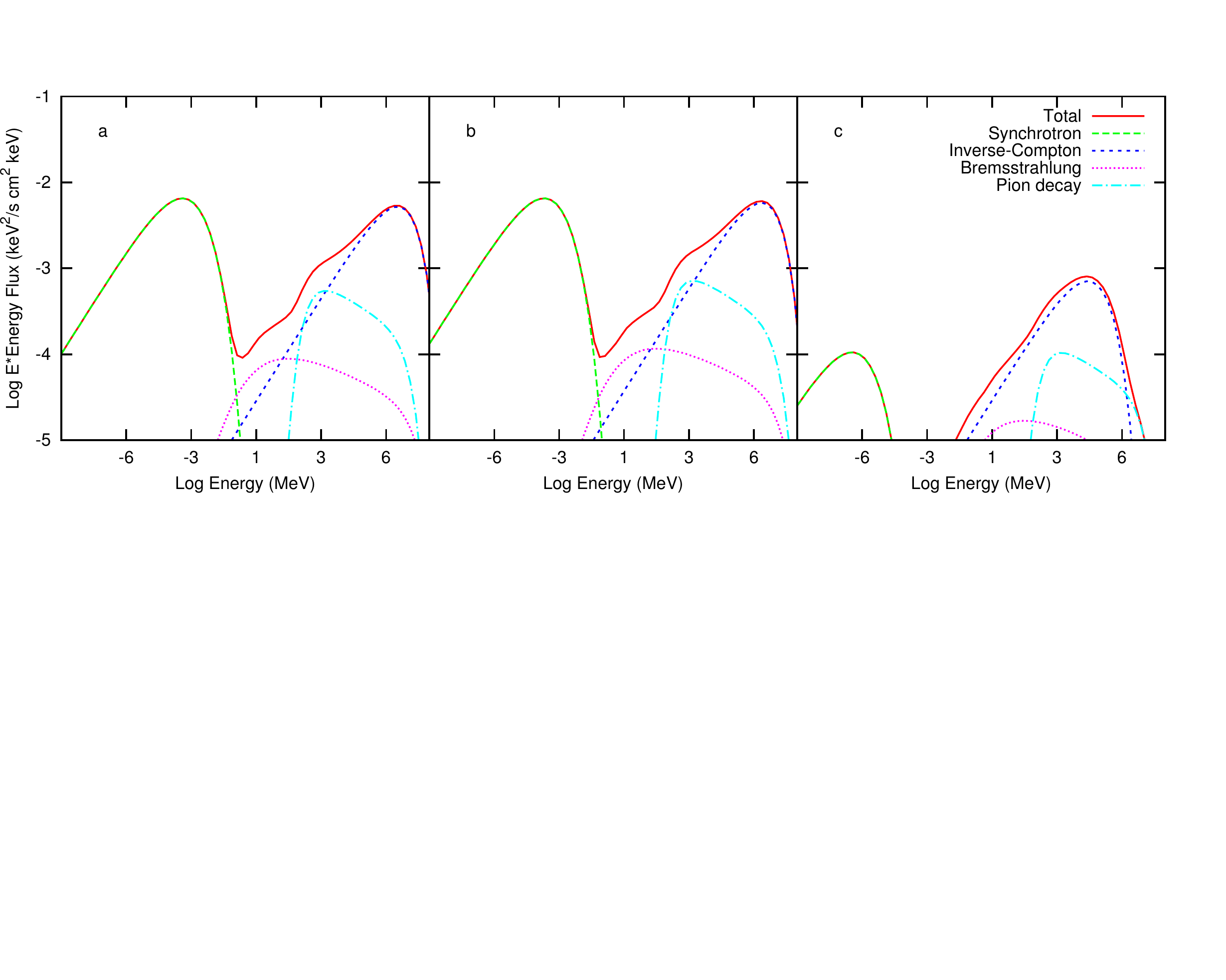}}
\caption{Spectra of models U1 (a), U2 (b), and U3 (c)).  Only at
very late times, as exhibited in (c), is there a substantial
difference in the spectrum, as the much lower shock velocity 
greatly reduces the populations of accelerated ions and electrons.}
\label{unispecs}
\end{figure}

Figure~\ref{unifluxes} shows the time evolution of fluxes (shown as
$E^2 F(E)$ keV$^2$ cm$^{-2}$ s$^{-1}$ keV$^{-1}$, where $F(E)$ is the
  number flux of photons, or $E S(E)$ where $S(E)$ is the energy flux)
  due to the different radiative processes, for the case of expansion
  into a uniform medium.  While the dynamics are self-similar, the
  evolution of the electron distribution is not, and there is some
  structure to the flux evolution, though the variations are not
  large.  At late times, the slowing shock results in both a lower
  maximum energy for ions, and in fewer relativistic particles.
  Figure~\ref{unispecs} shows SEDs for the three models U1, U2, and
  U3.

\subsection{Collision with cavity wall}

\begin{figure}
\centerline{\includegraphics[scale=0.45]{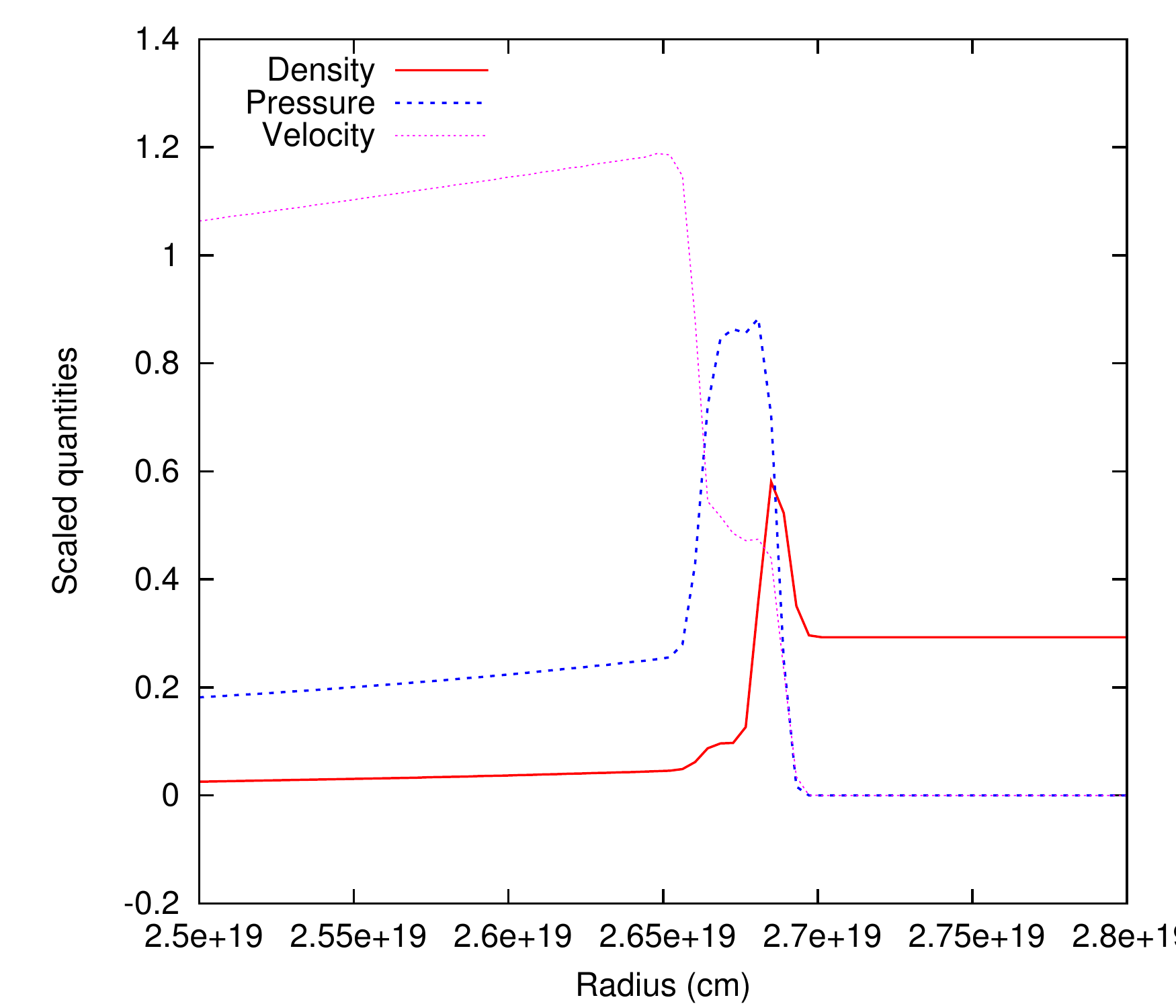}\hskip10truept
  \includegraphics[scale=0.45]{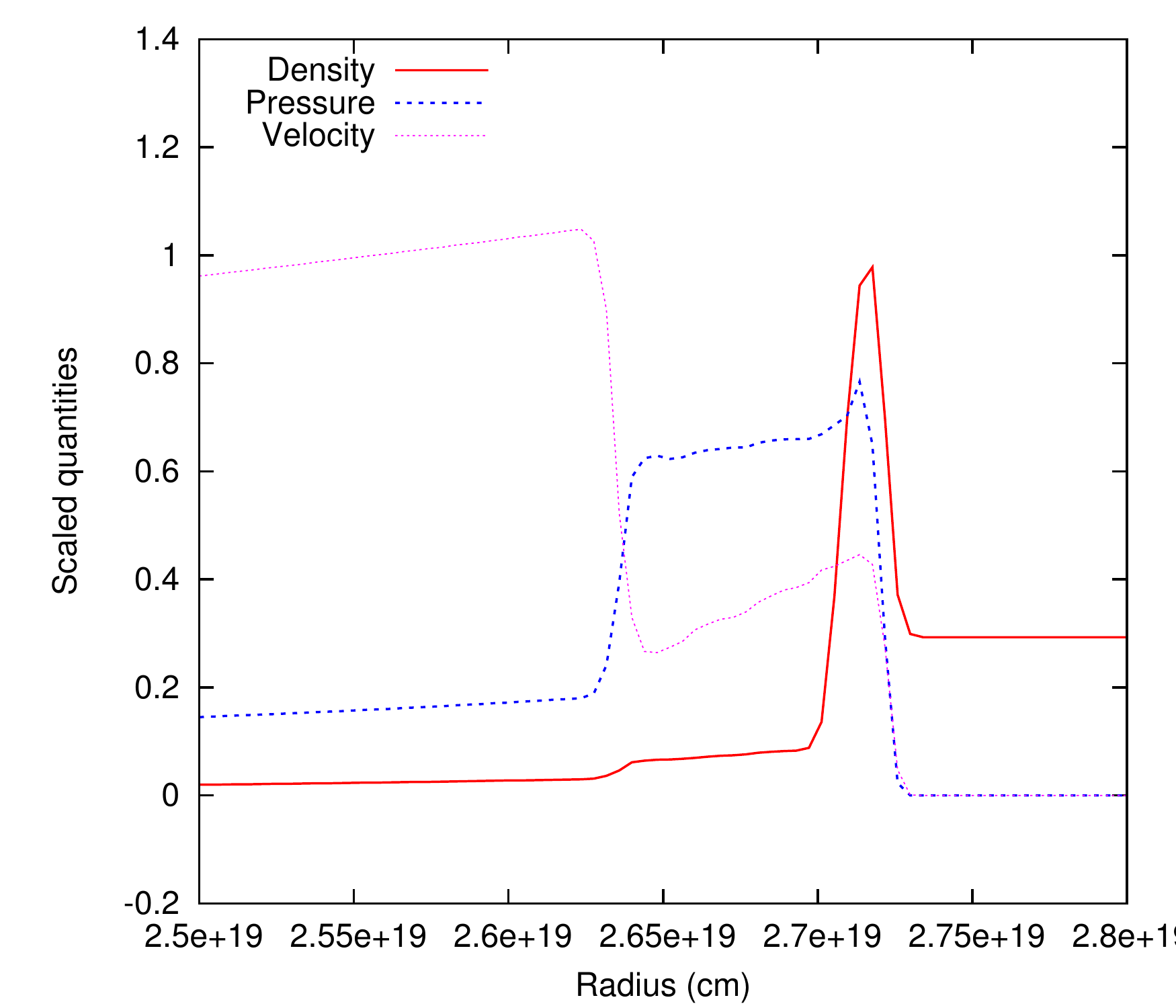}}
\centerline{\includegraphics[scale=0.45]{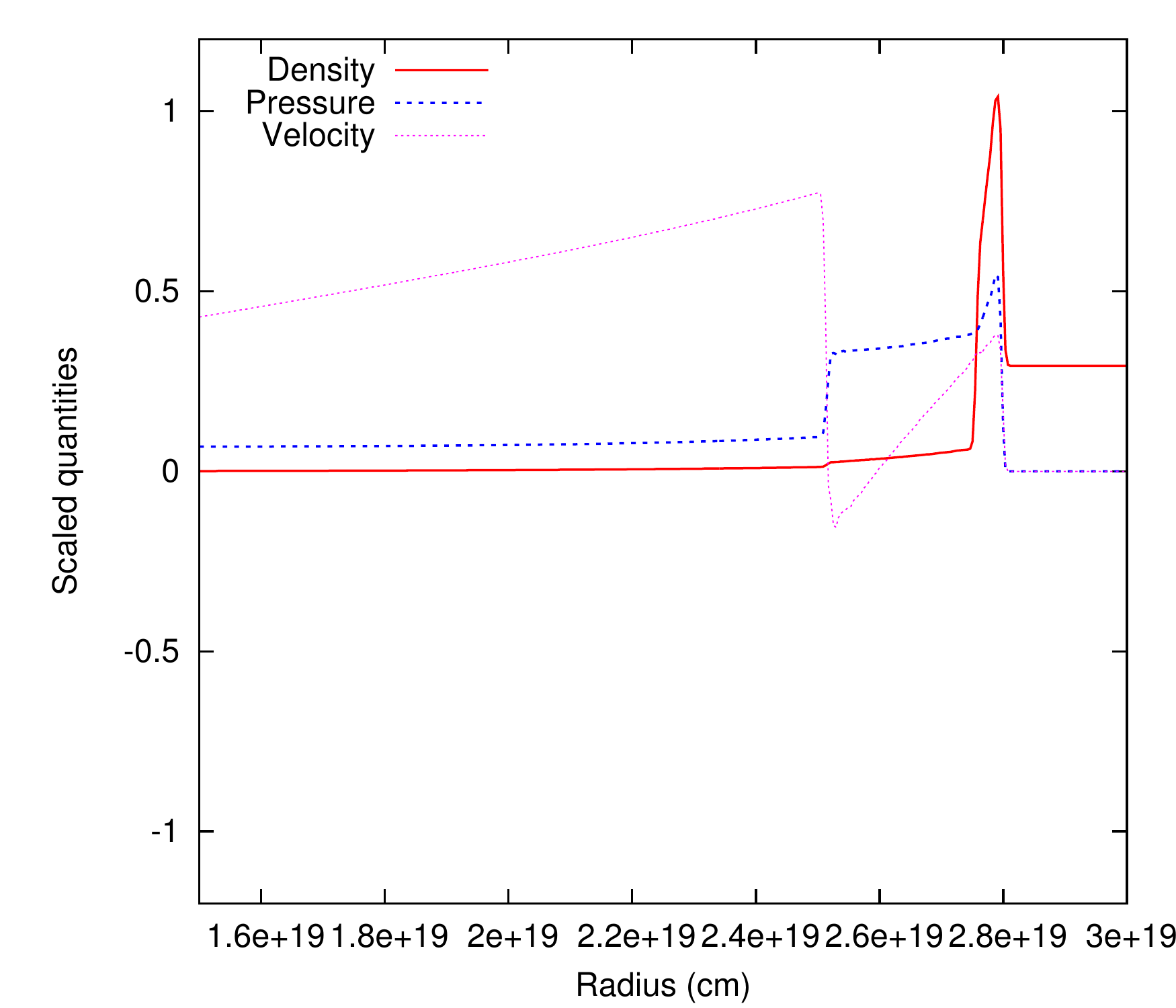}\hskip10truept
  \includegraphics[scale=0.45]{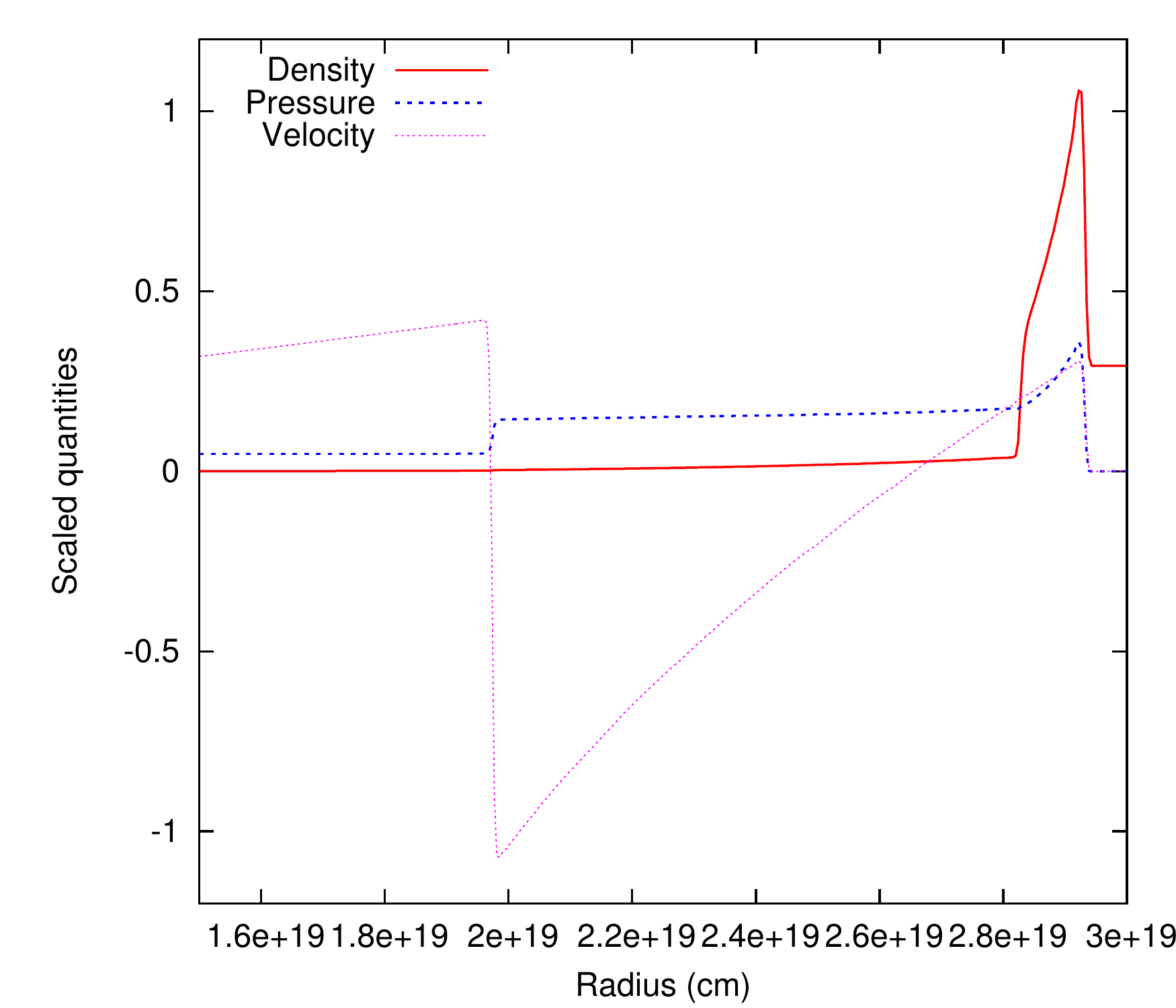}}
\caption{Four snapshots of the blast wave just after reaching the wall
  where the density jumps by a factor of 20.  The times are 2070,
  2210, 2650, and 3530 yr.  Densities are in units of $4 \times 10^{-23}$
cm$^{-3}$, pressures in units of $4 \times 10^{-8}$ dyn cm$^{-2}$, and
velocities in units of $1000$ km s$^{-1}$.  
Note the scale change between the third
  and fourth panels.  The reflected shock accelerates rapidly back
  toward the remnant center.}
\label{collision}
\end{figure}

For the standard collision model parameters, the collision with the
cavity wall occurs at a time of 1990 yr.  Figure~\ref{collision} shows
profiles of density, pressure, and velocity just before and just after
the collision with the wall.  The reflected shock (negative
velocities) moves back into the shocked material, while the blast wave
moves forward at a much lower velocity, roughly lower as the square
root of the higher density.

\subsubsection{Time evolution of spectra}

\begin{figure}
\centerline{\includegraphics[scale=0.8]{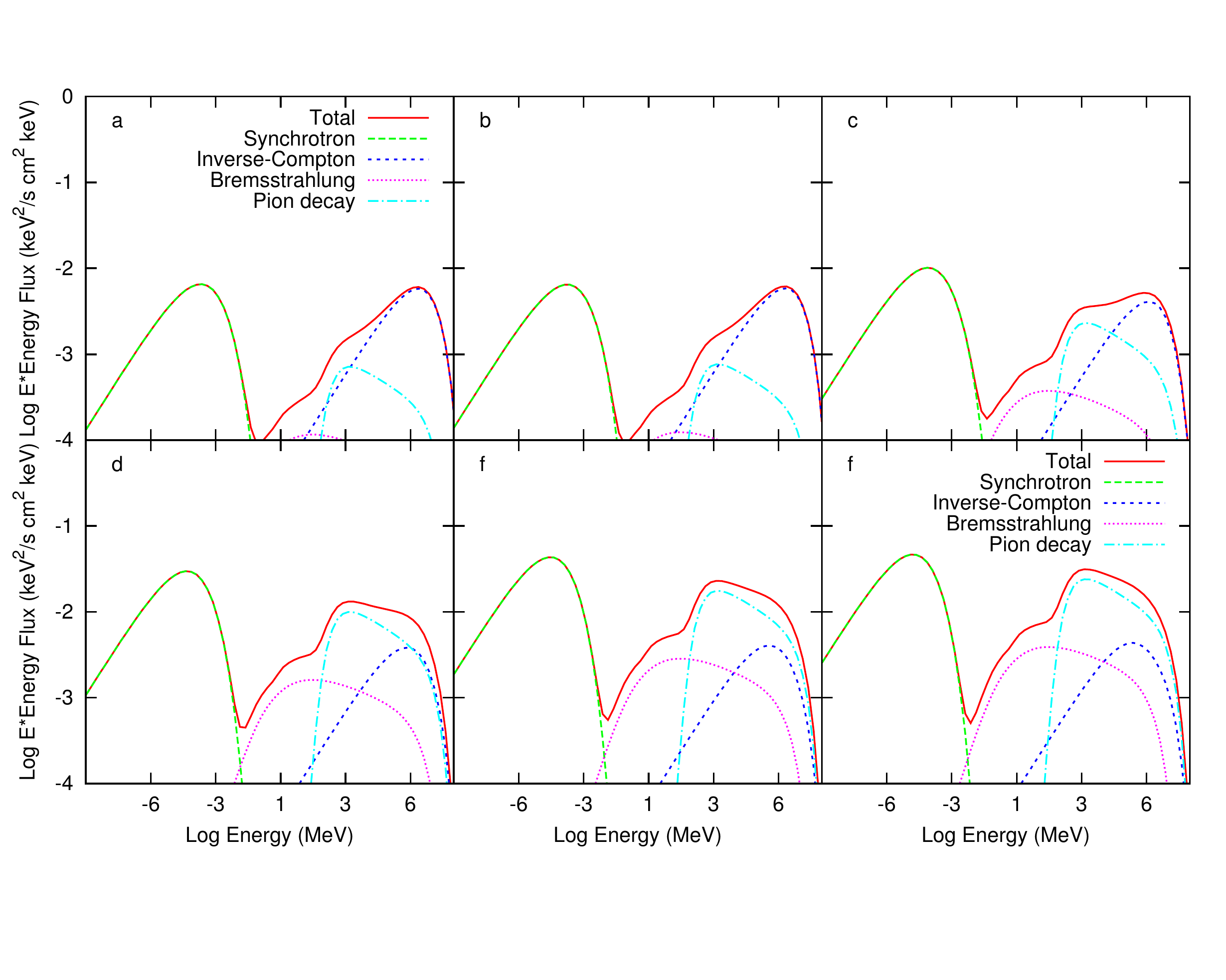}}
\caption{Spectra as a function of time, for the standard collision
model.  (a) JB (just before; age 1700 yr); (b) model A1 (2027 yr);
(c) model A2 (2236 yr); (d) model A3 (2740 yr);
(e) model A4 (3541 yr); (f) model A5 (5350 yr).  Note the change in
relative strengths of IC and $\pi^0$-decay processes with time, causing
dramatic changes in the summed spectrum at GeV to TeV energies.}
\label{scspectra}
\end{figure} 

Figure~\ref{scspectra} shows the individual spectra at the six times
of the standard-collision model listed in Table~\ref{scparameters}.
The synchrotron peak strengthens and moves to lower energies as the
shock velocity decreases.  However, the major change is in the shape
of the GeV-TeV emission, where the increasing relative prominence of
$\pi^0$ and bremsstrahlung emission relative to inverse-Compton
emission causes the shape of the total spectrum to change dramatically
with time, from a positive slope (in $E^2 F(E)$ space) to a steeply
negative slope.  Since the former two processes basically vary as
density squared while the latter only as the first power of density,
the much greater density in the wall is responsible for this change.

The SEDs after the collision show significant differences from the JB
model shown in Fig.~\ref{sedovannum}, mainly in the gamma-ray regime.
The much greater density for targets increases the importance of
$\pi^0$ emission and bremsstrahlung relative to ICCMB, and the sum
actually mimics a power-law with a softer spectrum than that of
$\pi^0$ emission alone.  At a significantly later time, model A5 shows
the effects of a slower shock velocity and greater volume of shocked
denser material.  The slower shock has reduced the maximum electron
energy, since it is limited by radiative losses. The maximum ion
energy continues to rise, but only slowly, with the increased age
almost completely canceled by the slowing shock.  The total emission
has a significantly steeper slope between 1 GeV and 10 TeV than in
model A2, closer to that of $\pi^0$-decay alone (but not yet as
steep).

\subsubsection{Images}

\begin{figure}
\centerline{Model A1 (age 2027 yr)}
\vspace{0.3truein}
\centerline{\includegraphics[scale=0.35]{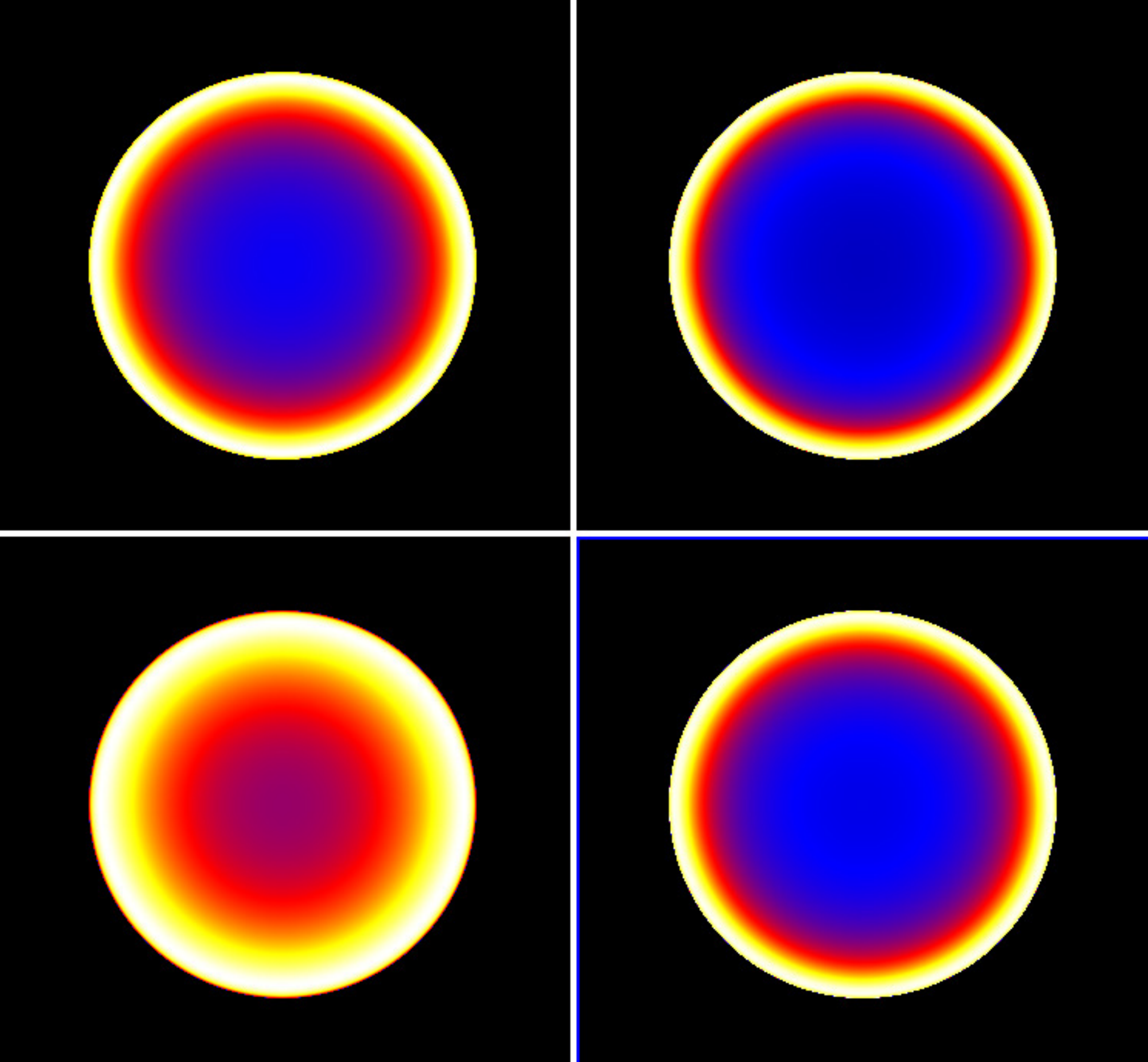}\hskip10truept
  \includegraphics[scale=0.4]{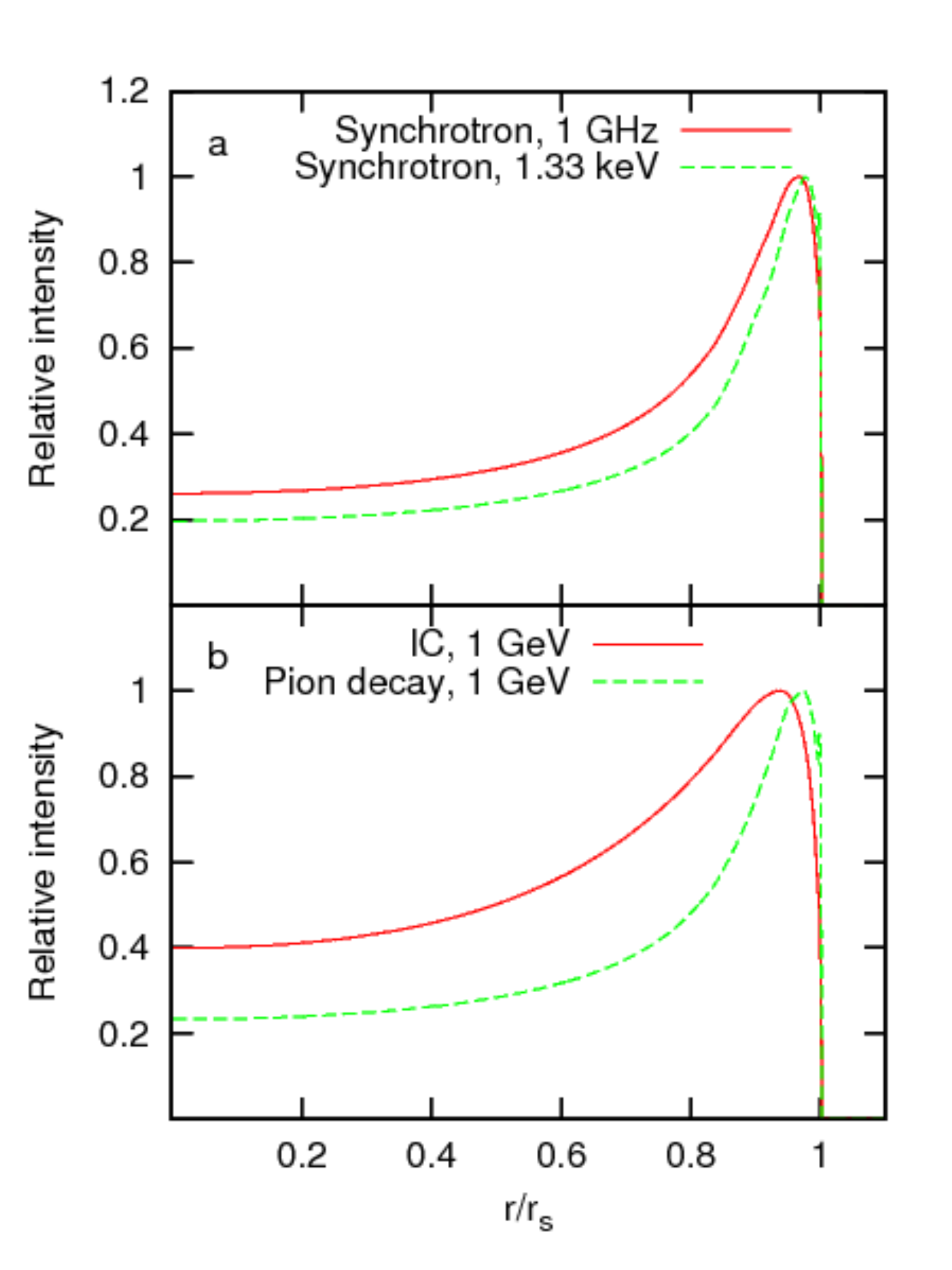}\hskip10truept}
\vskip10truept
%\centerline{\includegraphics[scale=0.35]{a7-6ims.eps}\hskip10truept}
\caption{Images and profiles due to different radiative processes, for
  model A1.  Top row: Synchrotron emission at 1 GHz and 1.33 keV, and
  profiles (a).  Bottom row: IC and $\pi^0$ emission at 1 GeV, and
  profiles (b).  The images and profiles for IC and $\pi^0$ emission
  at 1 TeV are indistinguishable from those at 1 GeV, and the
  bremsstrahlung image is very similar to the $\pi^0$-decay image
  throughout.  The color scale is relative to each image maximum.}
\label{a737-4ims}
\end{figure}

\begin{figure}
\centerline{Model A2 (age 2236 yr)}
\vspace{0.3truein}
\centerline{\includegraphics[scale=0.37]{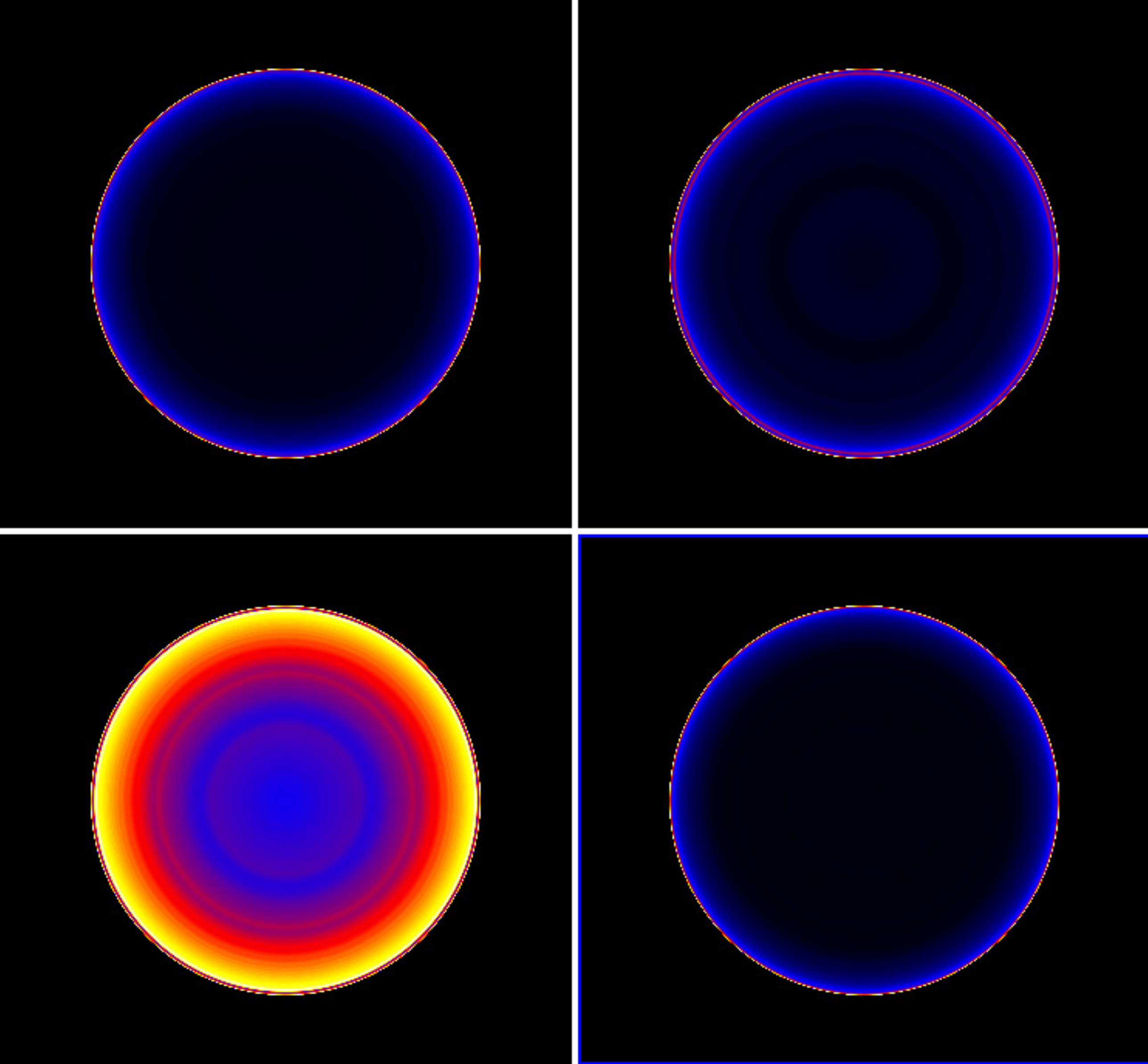}\hskip10truept
  \includegraphics[scale=0.42]{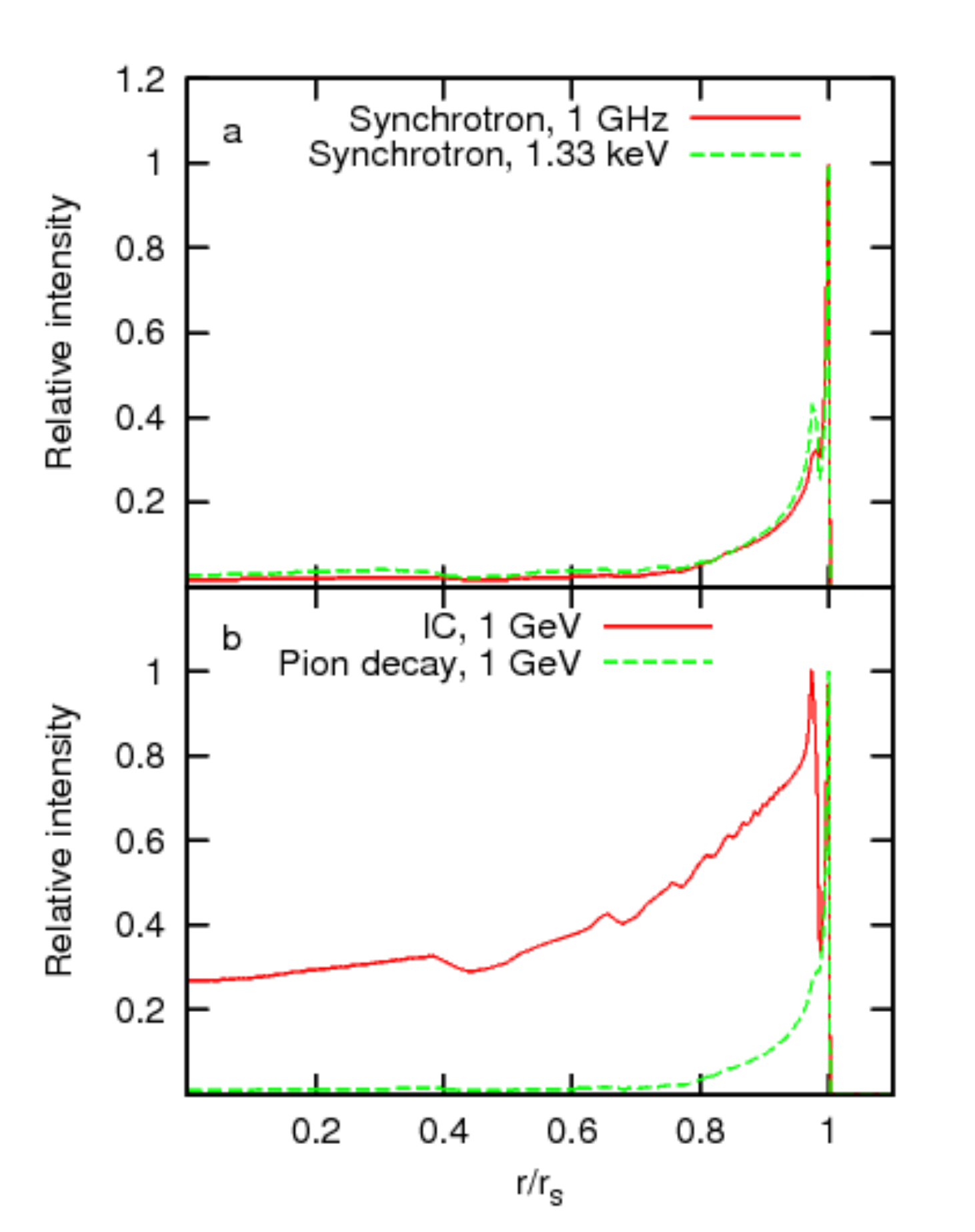}\hskip10truept}
\vskip10truept
%\centerline{\includegraphics[scale=0.35]{a7-6ims.eps}\hskip10truept}
\caption{Images and profiles due to different radiative processes, for
  model A2.  Top row: Synchrotron emission at 1 GHz and 1.33 keV, and
  profiles (a).  Bottom row: IC and $\pi^0$ emission at 1 GeV, and
  profiles (b) (indistinguishable from those at 1 TeV).  The small
  ripples in the IC image are artifacts of the numerical
  interpolations.  Again, the color scale is relative to each image
  maximum.}
\label{a586-4ims}
\end{figure}

\begin{figure}
\centerline{Model A3 (age 2740 yr)}
\vspace{0.3truein}
\centerline{\includegraphics[scale=0.37]{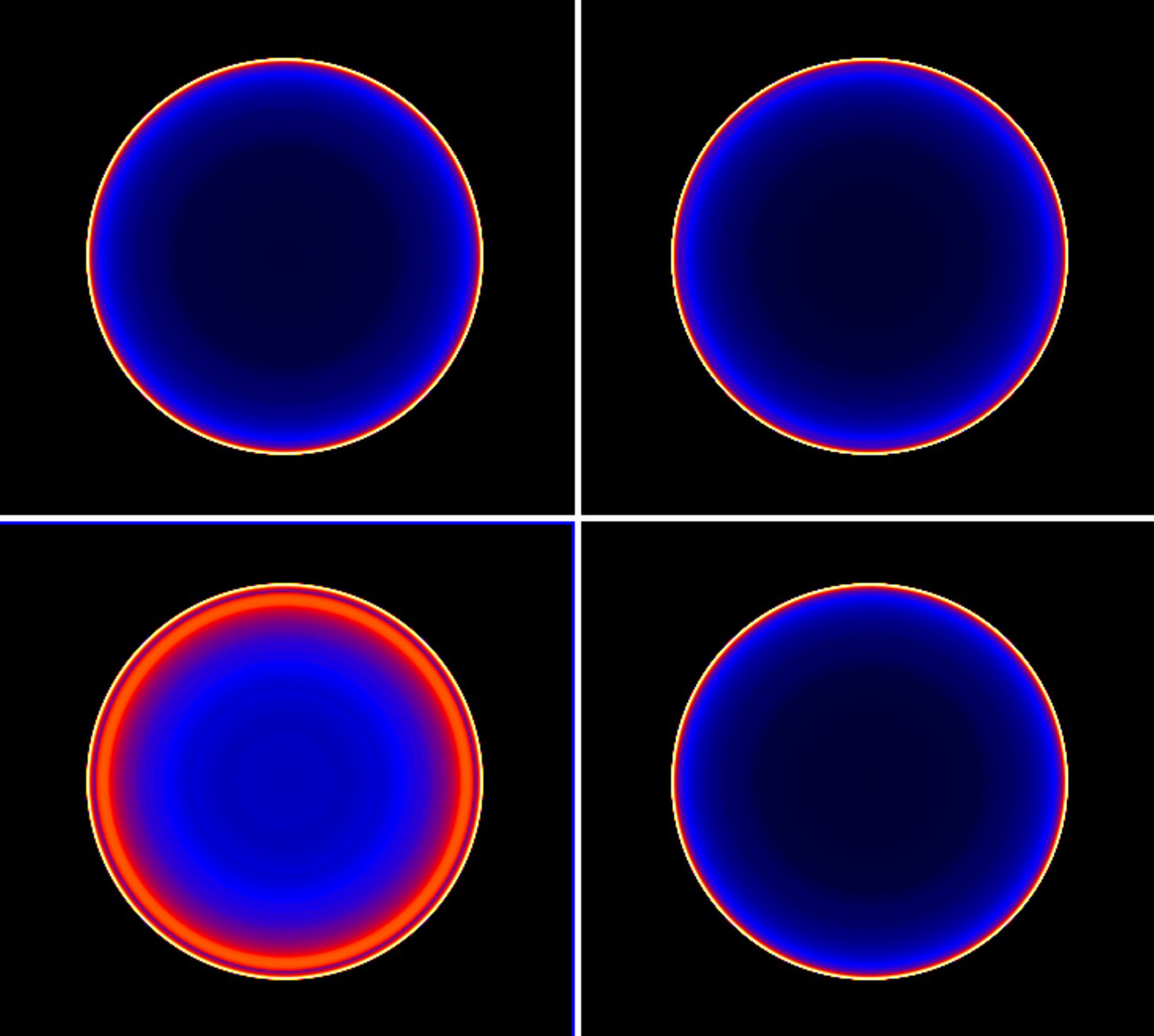}\hskip10truept
  \includegraphics[scale=0.42]{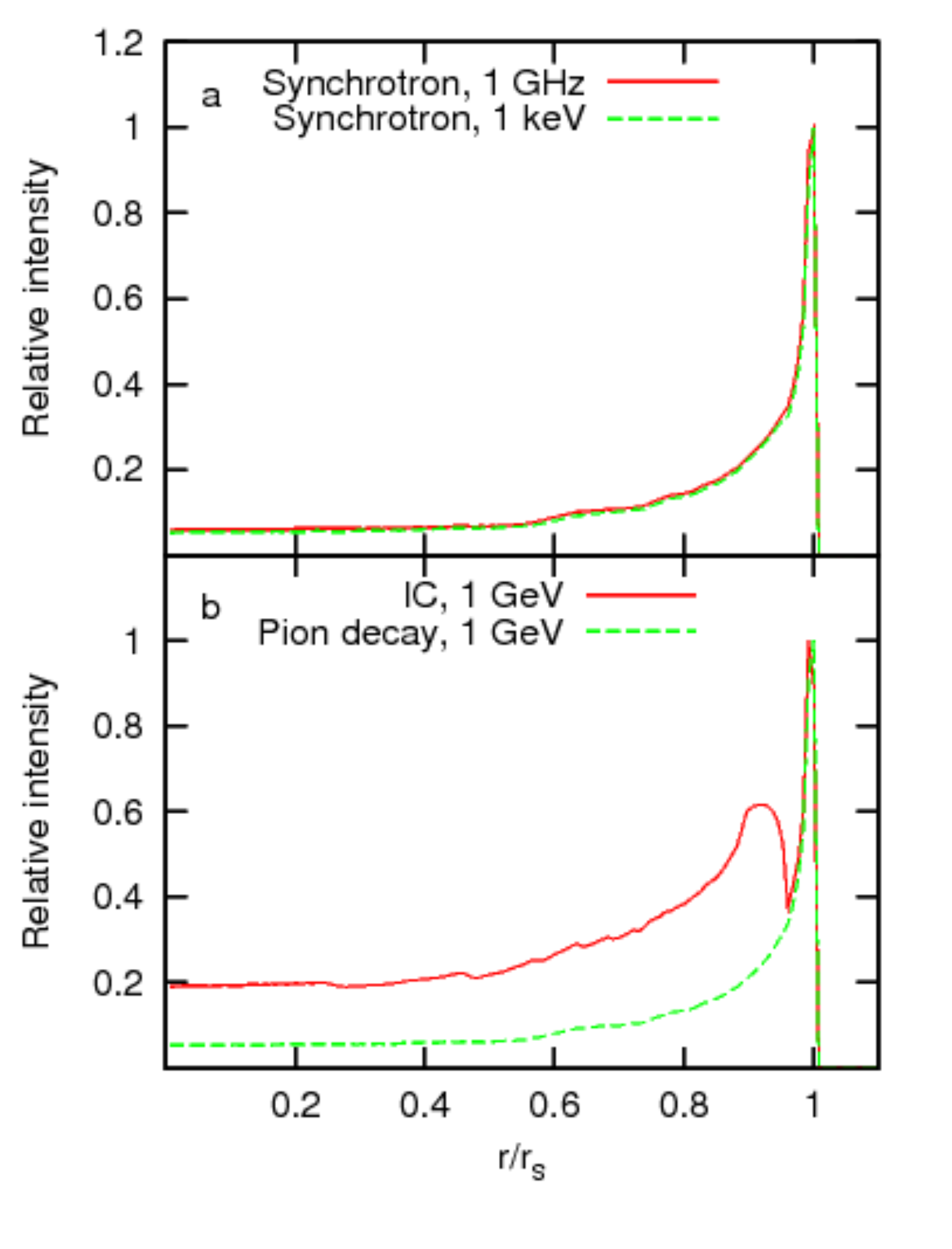}\hskip10truept}
\vskip10truept
%\centerline{\includegraphics[scale=0.35]{a7-6ims.eps}\hskip10truept}
\caption{Images and profiles due to different radiative processes, for model
A3.  Top row: Synchrotron emission at 1 GHz and 1.33 keV, and profiles (a).
  Bottom row: IC and $\pi^0$ emission at 1 GeV, and profiles (b).}
\label{ja-4ims}
\end{figure}

\begin{figure}
\centerline{Model A4 (age 3541 yr)}
\vspace{0.3truein}
\centerline{\includegraphics[scale=0.5]{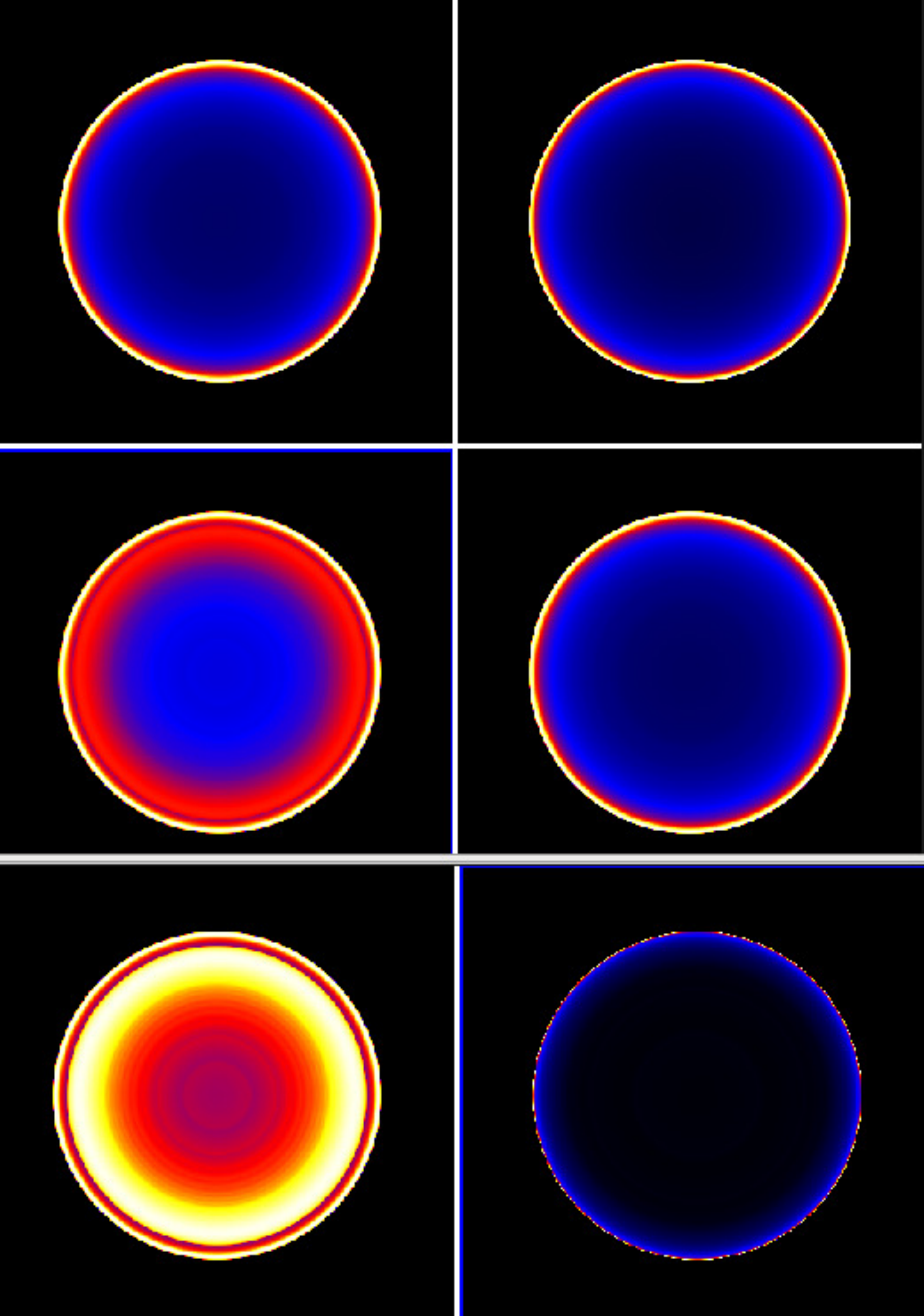}\hskip10truept
  \includegraphics[scale=0.64]{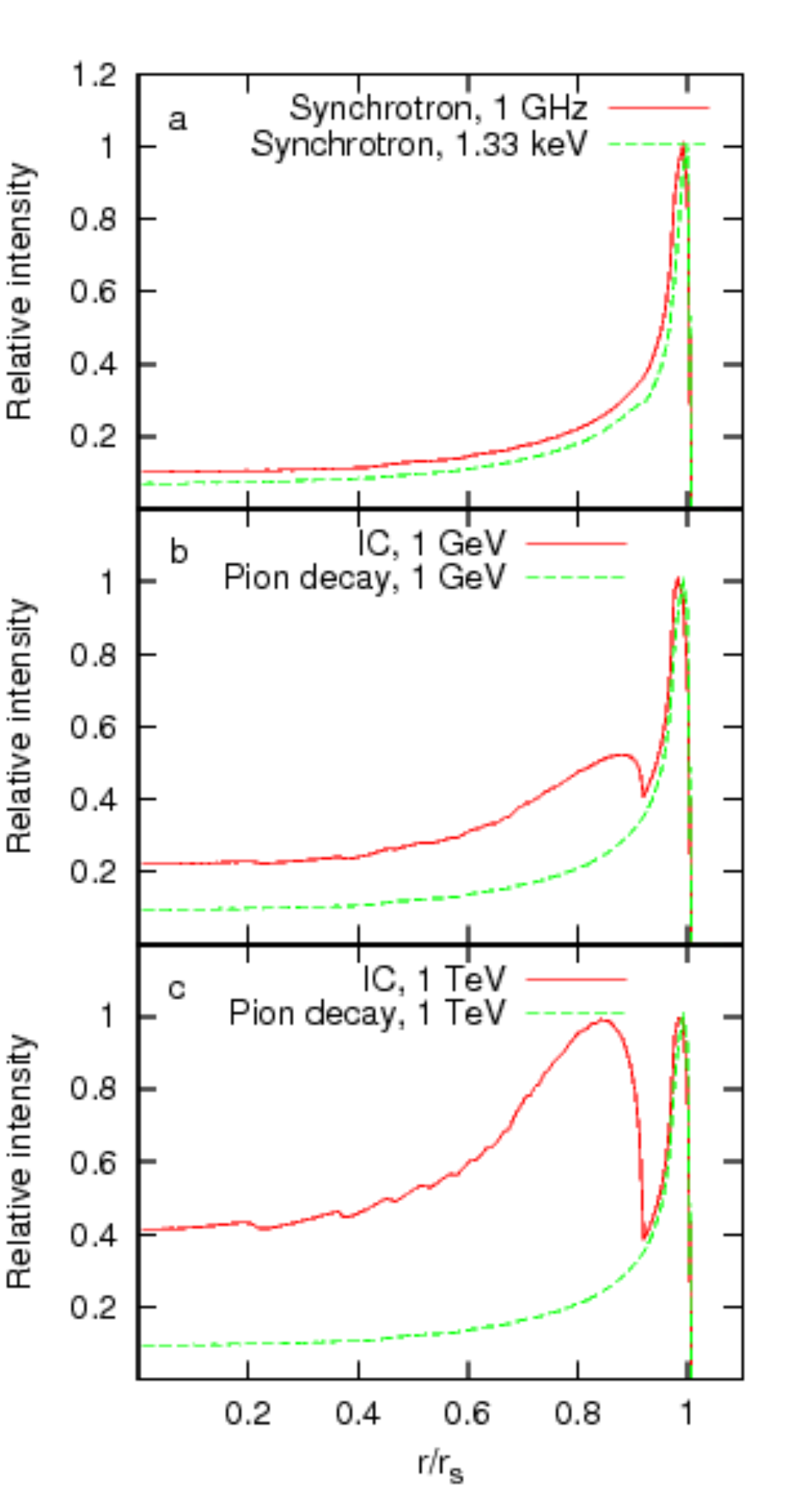}\hskip10truept}
\vskip10truept
%\centerline{\includegraphics[scale=0.35]{a7-6ims.eps}\hskip10truept}
\caption{Images and profiles due to different radiative processes, for model
A4.  Top row: Synchrotron emission at 1 GHz and 1.33 keV, and profiles (a).
  Middle row: IC and $\pi^0$ emission at 1 GeV, and profiles (b).
  Bottom row: IC and $\pi^0$ emission at 1 TeV, and profiles (c).}
\label{a413-6ims}
\end{figure}

\begin{figure}
\centerline{Model A5 (age 5350 yr)}
\vspace{0.3truein}
\centerline{\includegraphics[scale=0.4]{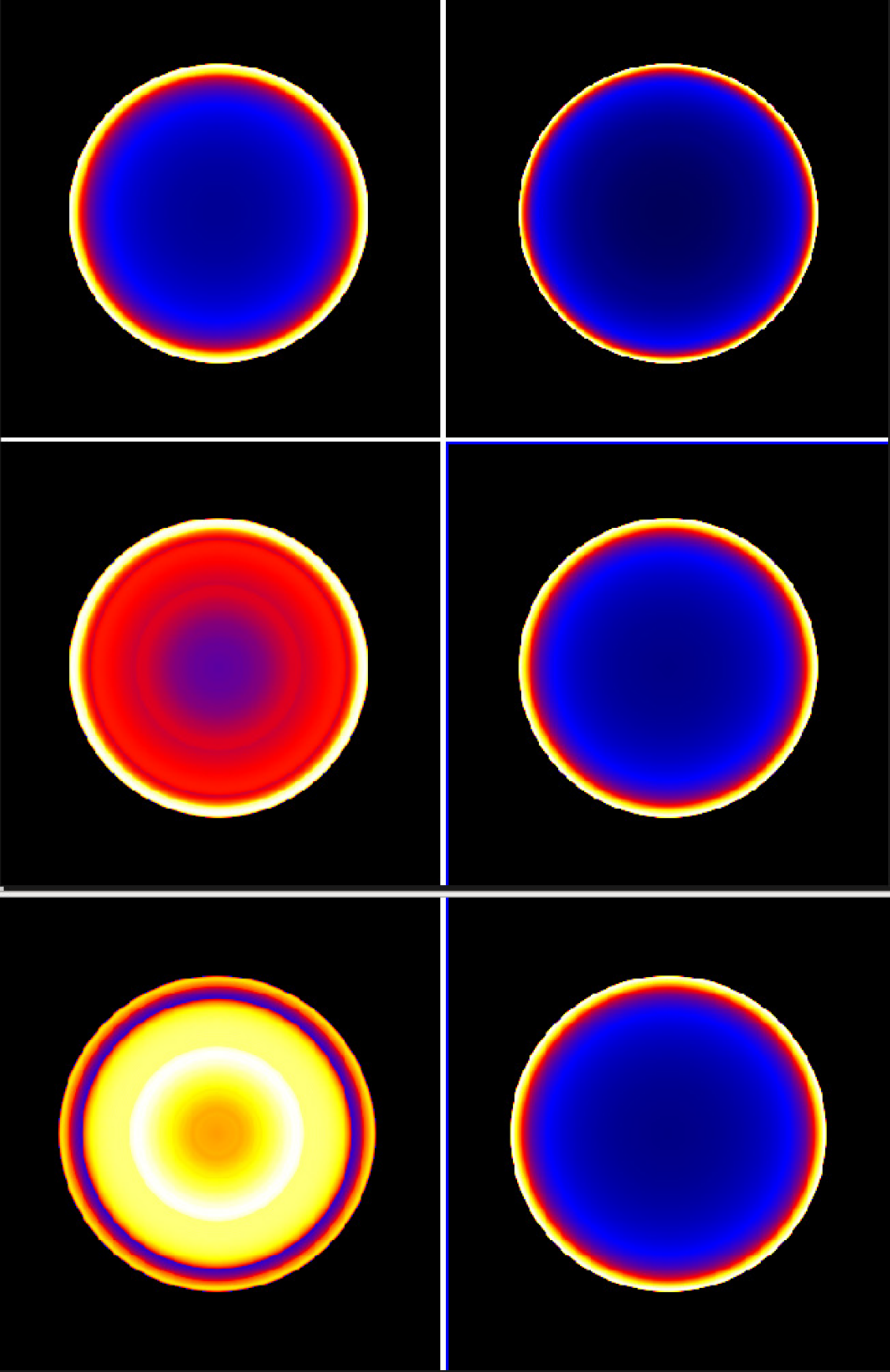}\hskip10truept
  \includegraphics[scale=0.5]{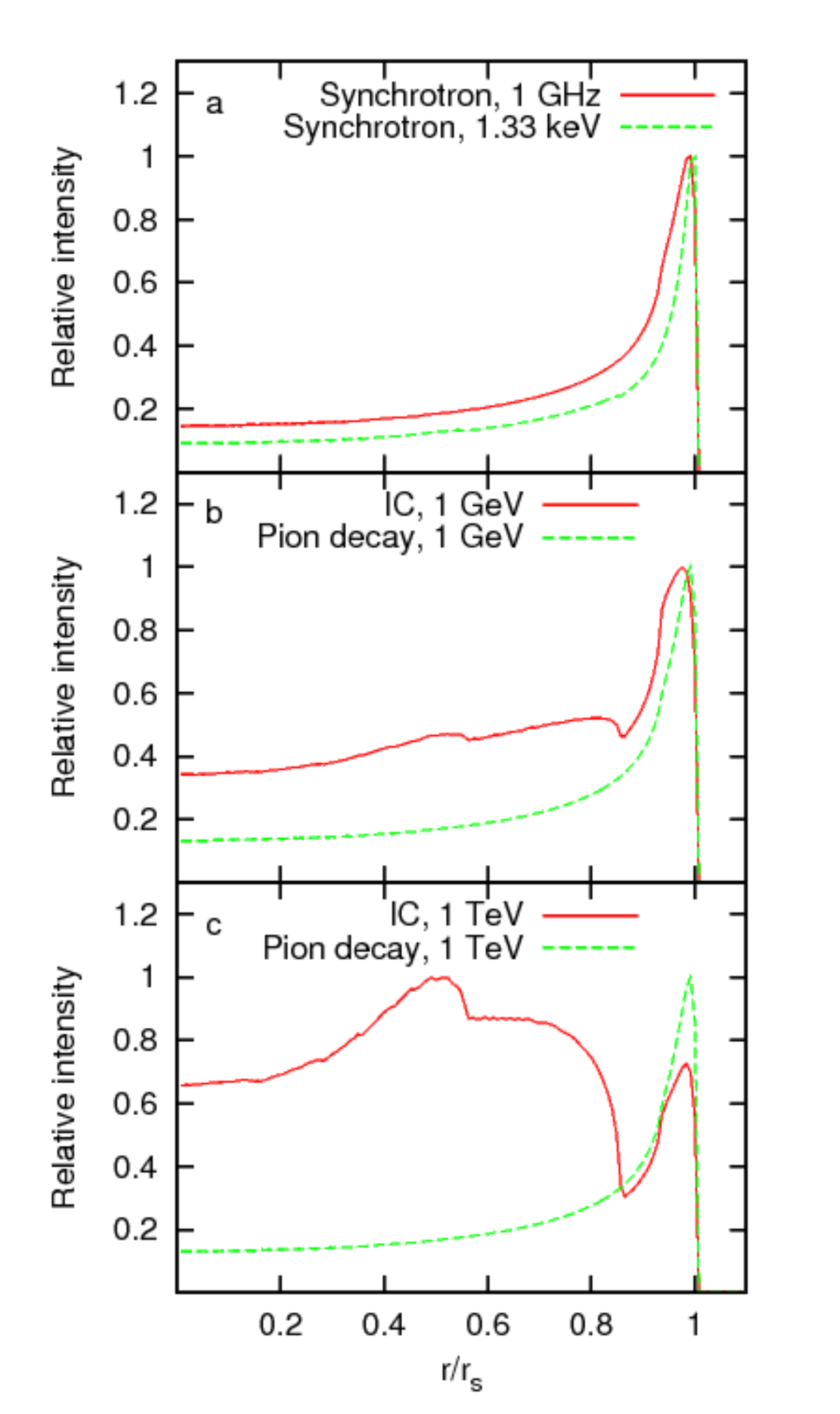}\hskip10truept}
\vskip10truept
%\centerline{\includegraphics[scale=0.35]{a7-6ims.eps}\hskip10truept}
\caption{Images and profiles due to different radiative processes, for model
A5.  Top row: Synchrotron emission at 1 GHz and 1.33 keV, and profiles (a).
  Middle row: IC and $\pi^0$ emission at 1 GeV, and profiles (b).
  Bottom row: IC and $\pi^0$ emission at 1 TeV, and profiles (c).
The bump near $r/r_s = 0.5$ for IC emission at 1 TeV is an artifact of the
numerical interpolations, visible only at the highest energies.}
\label{a5-6ims}
\end{figure}

Figs.~\ref{a737-4ims} -- \ref{a5-6ims} show model images and profiles.
In model A1 (Fig.~\ref{a737-4ims}), the
emission is dominated by material in the cavity; the shocked wall
region is still very thin.  The X-ray synchrotron emission is confined
to a thinner shell than the radio emission, due to electron energy
losses.  The IC emission occupies more of the remnant interior because
of the dependence on only one power of density.  The images at 1 GeV
and 1 TeV are very similar and the latter are not shown.  The next two
sets of images, for models A2 and A3, show the effects of the rapidly
increasing emission from shocked wall material.  The synchrotron
emission is almost totally confined to the region containing shocked
wall material, but that emission is much brighter than before; the
emission from shocked cavity material is much less prominent by
contrast.  The $\pi^0$ emission (and bremsstrahlung, not shown but
indistinguishable from the $\pi^0$ morphology) are similarly confined
to a thin rim, since their emissivities also depend on the square of
density.  However, the inverse-Compton emission, which varies only
with the first power of density, has a much stronger presence in the
remnant interior.  In the later models A4 and A5
(Figs.~\ref{a413-6ims} and \ref{a5-6ims}), there are substantial
differences in the IC emission between 1 GeV and 1 TeV; at 1 TeV the
IC emission actually peaks at the interface between shocked wall
material and shocked cavity material, instead of at the outer blast
wave, and has a much larger center-to-limb ratio in brightness.  The
profiles of Figs.~\ref{a737-4ims} -- \ref{a5-6ims} show these effects
more clearly.

\subsection{Time evolution of fluxes}

\begin{figure}
\centerline{\includegraphics[scale=0.4]{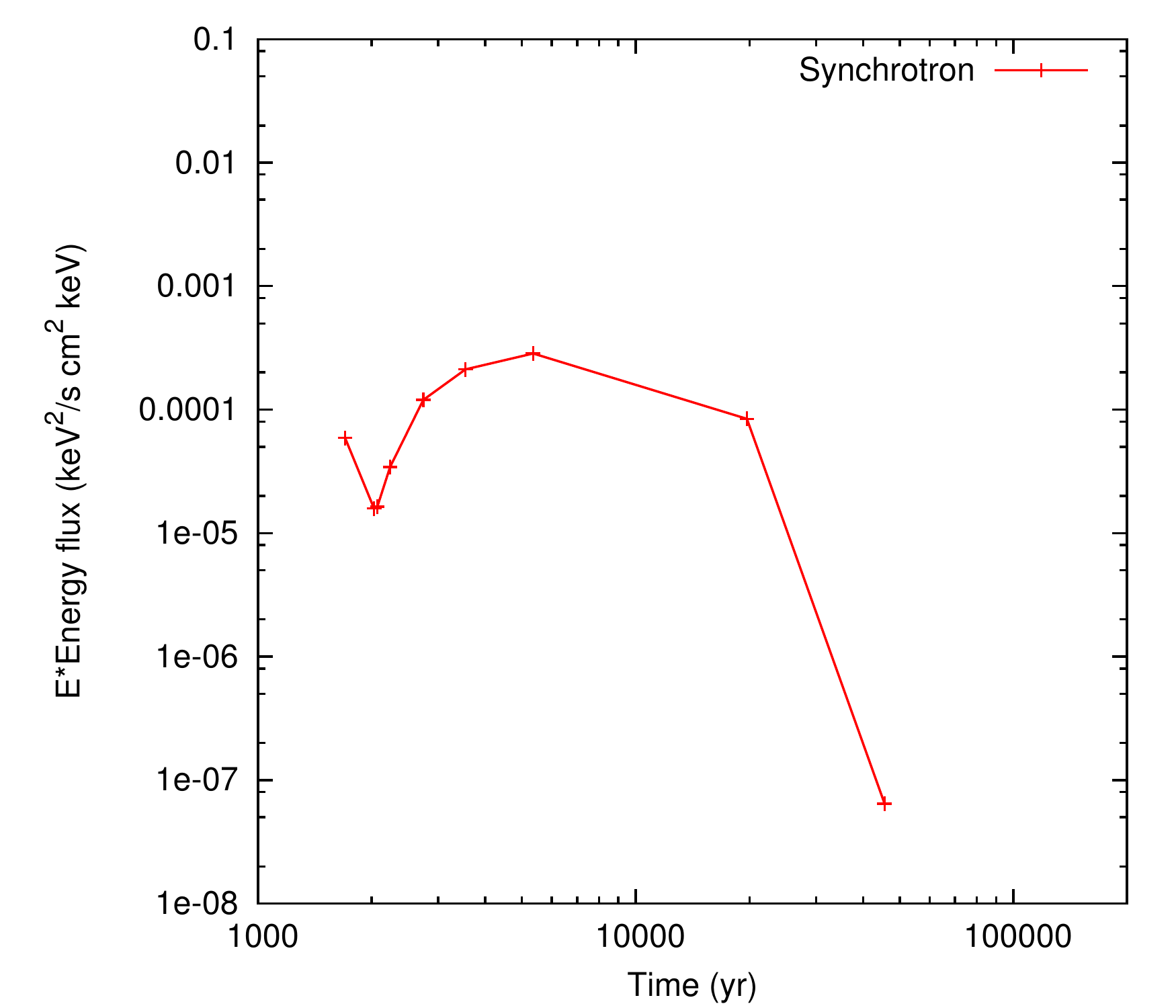}\hskip10truept
  \includegraphics[scale=0.4]{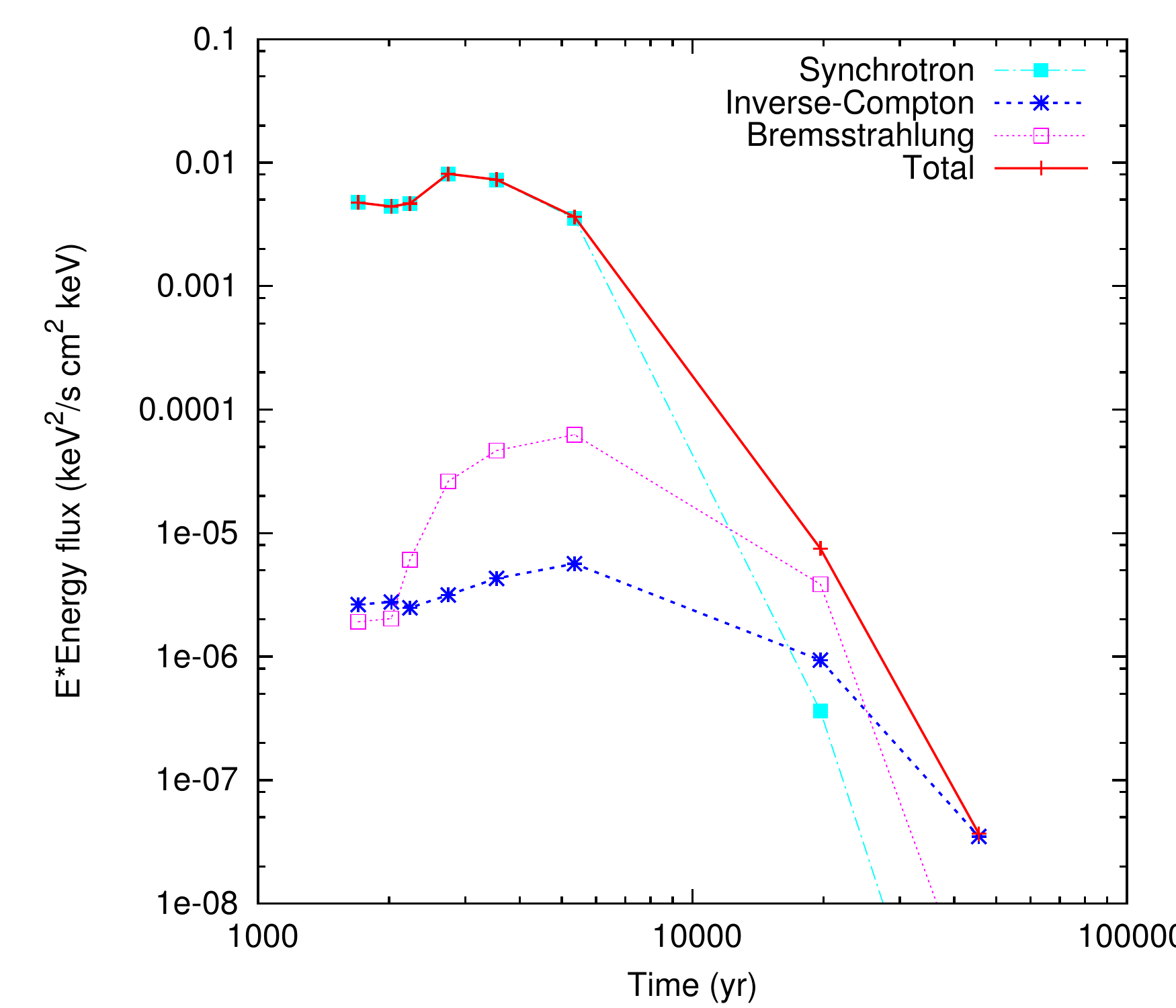}}
\centerline{\includegraphics[scale=0.4]{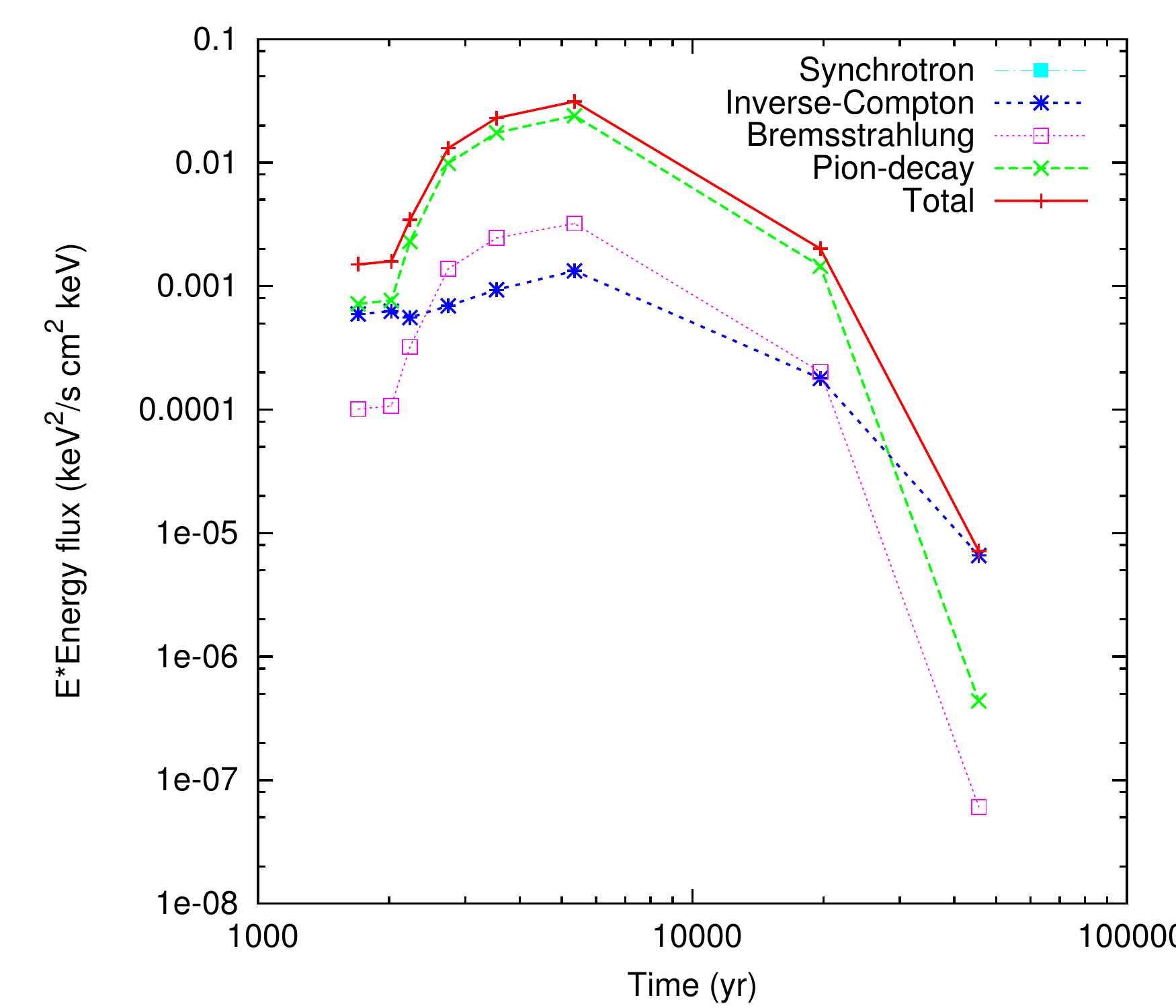}\hskip10truept
  \includegraphics[scale=0.4]{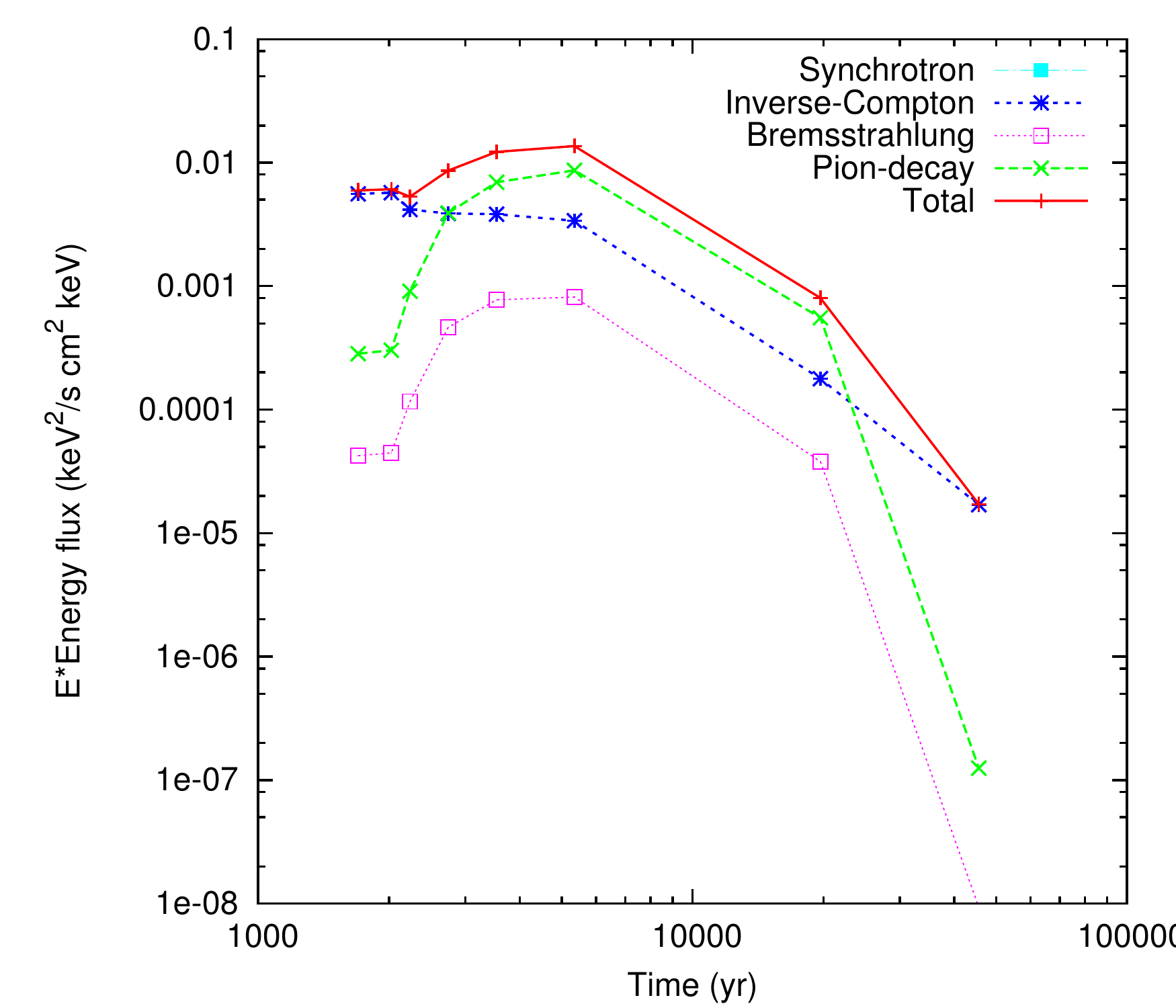}}
\caption{Fluxes as a function of time, for the standard collision
model.  Upper left:  1 GHz.  Upper right: 1.33 keV.  Lower left:  1 GeV.
Lower right:  1 TeV.}
\label{scbfluxes}
\end{figure}

For the standard-collision case, the time variation of fluxes is of
course quite different from the uniform-medium case.  The fluxes due
to the different processes as a function of time are shown in
Fig.~\ref{scbfluxes}.  The time of the collision is 1990 yr, just
after the first datapoint.  Fluxes rise rapidly thereafter for a few
thousand years, eventually resuming the decline due to expansion and
slowing shock velocity.  At the latest times, the IC process has the
slowest decline rate and comes to dominate the integrated spectrum, as
also shown in the spectra (Fig.~\ref{scspectra}).  At those times, 
the synchrotron peak has moved to low enough energies that there is
negligible X-ray synchrotron emission, though the gamma-ray emission is
still quite significant.

%\subsection{Dependence on parameters}

\section{Discussion}

The comparison of one-zone to 1-D models shows significant
differences, and not just in the obvious shell morphology of the
latter.  Even in the uniform-medium case, variations in integrated
fluxes of factors of several occur, as described above.  However, with
the interaction with the cavity wall, much greater differences appear.
In morphology, the four processes fall into three classes: (1)
synchrotron, (2) bremsstrahlung and $\pi^0$-decay, and (3) IC.
Differences betweeen the first two are slight, since the emissivity of
synchrotron emission varies approximately as density squared, as do
those of bremsstrahlung and $\pi^0$-decay.  IC from CMB photons,
however, depends on only one power of density.  Thus, as the remnant
encounters the cavity wall, emission from the cavity material becomes
essentially negligible for all but IC, but the latter shows enough
central emission that a double-shell structure can even emerge
(Figs.~\ref{a413-6ims} and \ref{a5-6ims}).  Note that unless the
self-generated radiation field of the remnant is extremely intense,
this would still be true even in the presence of IC from optical-IR
seed photon fields in addition to the CMB, as those radiation energy
densities are not expected to vary much over the remnant volume.

\begin{figure}
\centerline{\includegraphics[scale=0.9]{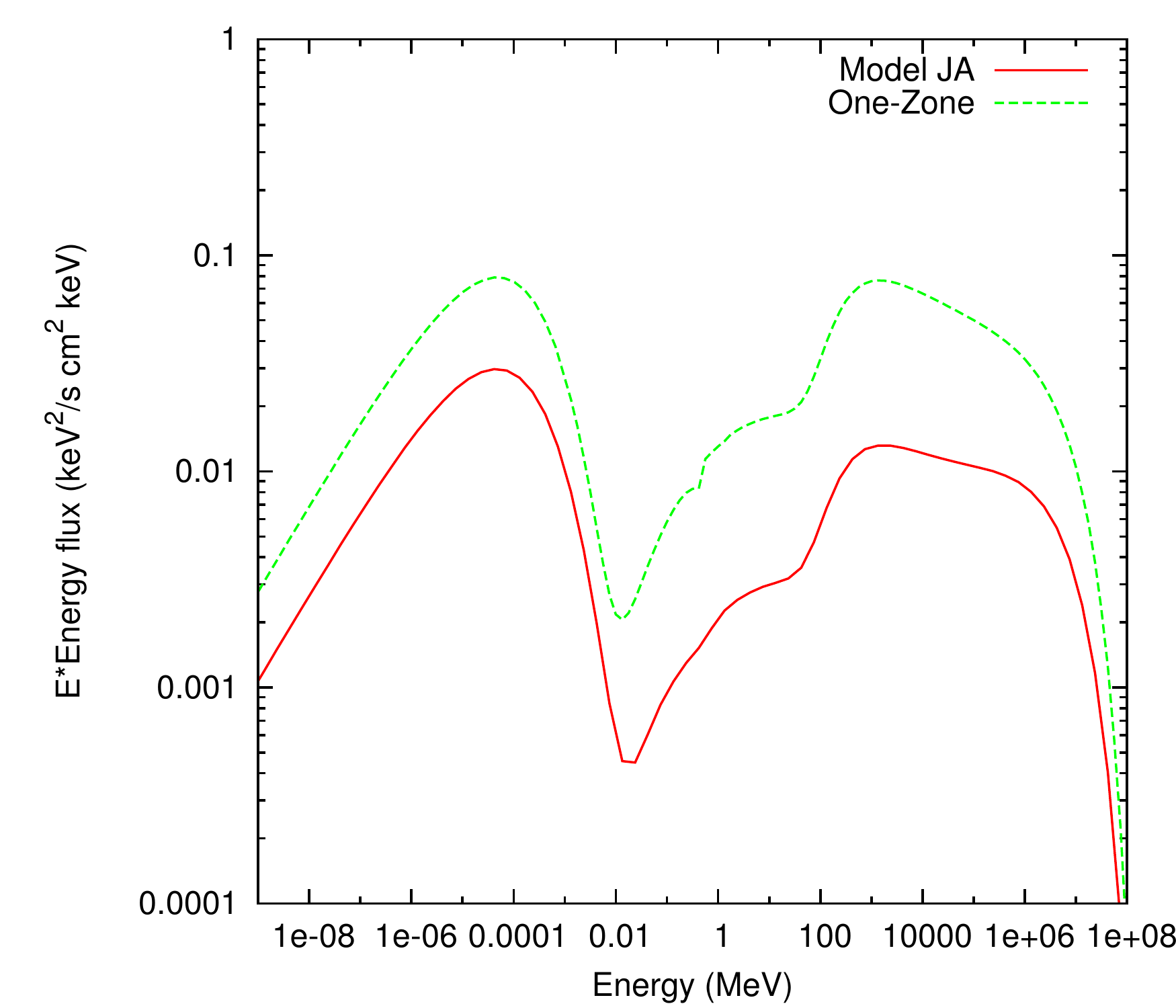}}
\caption{Comparison of the integrated spectrum of model A3 with
that from a homogeneous (one-zone) model using the parameters
of the forward blast wave (in the dense medium).}
\label{homvsja}
\end{figure}

This strong morphological difference has spectral consequences as
well.  Figure~\ref{homvsja} compares the emission after the collision
with the wall (model A3, age 2740 yr) with that from a homogeneous
(one-zone) model with the parameters of the current blast wave.  The
cavity in the past of model A3 substantially reduces the emission at
all wavelengths, but also changes the relative weights of the
different processes.  The higher relative weight of IC emission in
model A3 causes the slope of the GeV-TeV part of the spectrum to be
substantially flatter than for the one-zone model.  Clearly,
attempting to describe a radially varying source with the properties
of the current blast wave would result in incorrect inferences of
several parameters.

\begin{figure}
\centerline{\includegraphics[scale=0.4]{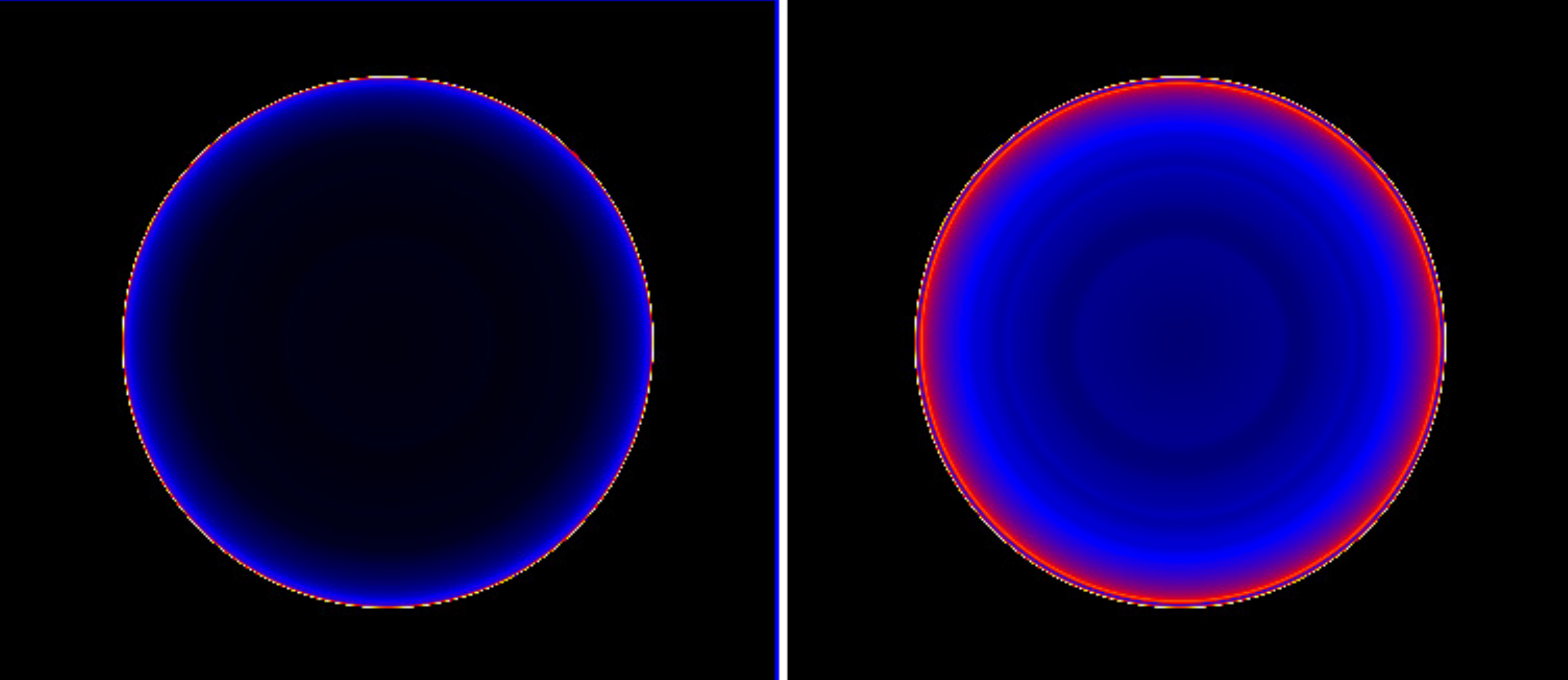}\hskip10truept
  \includegraphics[scale=0.32]{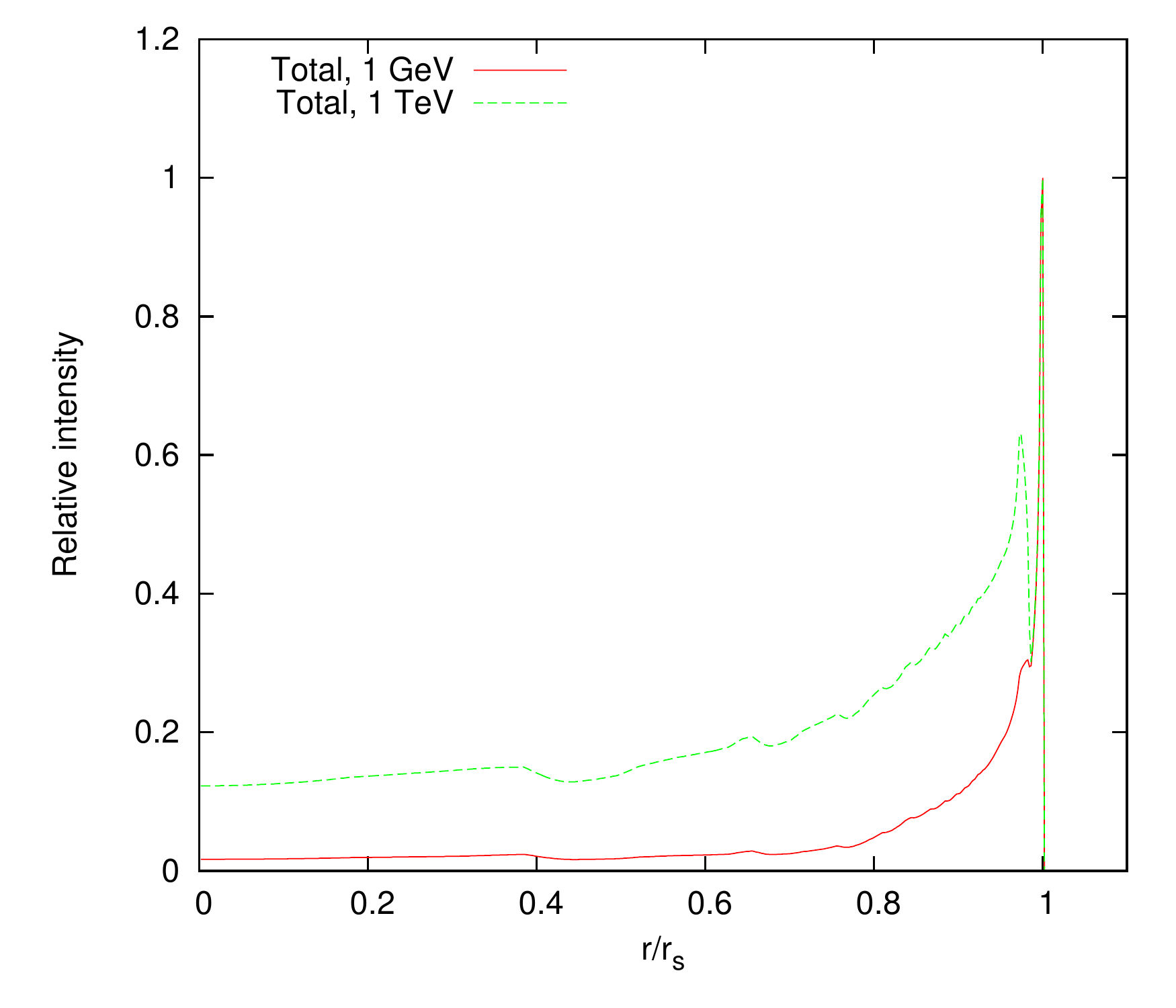}\hskip10truept}
\vskip10truept
\caption{Images of the total emission from model A2 (age 2236 yr) at
1 GeV (left) and 1 TeV (center), and profiles of each (right).
Images have $768 \times 768$ resolution.}
\label{a586-2ims}
\end{figure}

Figure~\ref{scspectra} shows that the total gamma-ray emission from
model A2 is dominated by $\pi^0$-decay at GeV energies and by IC at
TeV energies.  This implies also a strong morphological variation of
the source with energy.  Fig.~\ref{a586-2ims} shows the total emission
at 1 GeV and at 1 TeV, along with radial profiles of each.  While the
thin shell of shocked wall material is distinguishable in these
high-resolution calculations, such angular resolution will not be
available in this spectral regime for the foreseeable future.
However, even very coarse angular resolution can distinguish the much
greater central emission from the 1 TeV image, and such morphological
differences between GeV and TeV-range images are powerful clues to the
importance of hadronic vs.~leptonic processes in cavity SNRs.

\section{Summary and conclusions}

As angular resolutions and sensitivities of gamma-ray observations of
supernova remnants continue to improve, models will need to increase
in sophistication to allow appropriate interpretation of such new
observations.  We have systematically investigated the differences
between homogeneous (one-zone) models and those with radial variations
(spherical symmetry), for the remnant phases starting with times when
ejecta emission becomes unimportant, through the onset of radiative
cooling (Sedov blast waves, for power-law external media).  For the
case of a uniform external medium, analytic dynamics were used and
compared with results from hydrodynamic simulations.  For an explosion
in a constant-density cavity bounded by a wall of constant density 20
times larger, hydrodynamic simulations in spherical symmetry were
performed using the VH-1 code.  In both cases, the blast wave was
assumed to accelerate $10^{-4}$ of the ions with momenta greater than
ten times thermal (i.e., $10 m_p v_{\rm shock}$) to a power-law energy
distribution with power-law index 2.2; electrons were assumed to have
the same energy distribution with a normalization of 0.02 times that
of ions.  Both distributions were cut off exponentially at a maximum
energy limited by the remnant age, or for electrons, at a maximum
energy limited by radiative losses if lower.  Emission from these
distributions was calculated due to synchrotron radiation,
electron-ion and electron-electron bremsstrahlung, and inverse-Compton
upscattering of cosmic microwave background photons (ICCMB).  More
extensive sets of models, also showing the effects of variations of
parameters, are included in Tang (2016; PhD thesis, in preparation).

We find that for a uniform external medium, aside from obvious
morphological differences, one-zone models differ from 1-D spherical
models in integrated fluxes by factors of several.  Ratios of emission
due to different processes could differ by a somewhat larger factor.

We followed the time evolution of spectra, total fluxes, and remnant
morphology as the blast wave interacted with the cavity wall.  Here,
use of the current shock wave properties (shock speed and immediate
postshock density) in a one-zone model would produce substantially
different predictions compared to those of the cavity model.
Furthermore, the time evolution of the spectrum at GeV -- TeV energies
is such that apparent power-laws of different slope (and even
different sign, in $E^2 F(E)$ space) appear at different times, even
though the basic input spectrum does not change.  Interpretation of
slopes of power-law fits at gamma-ray energies should be undertaken
with great care; such slopes may not reflect the energy distribution
of a single underlying particle population.

Cavity-wall interactions produced striking morphological effects, from
significant variations of morphology with energy, to double-shell structures
which could potentially be resolvable.  Even with shock velocities as low
as a few hundred km s$^{-1}$, significant gamma-ray emission, diagnostic
of interactions with material with radially varying density, can occur in
older SNRs.  Such objects should be targets of future observations with
both current observatories such as {\sl Fermi} and H.E.S.S., and future
ones such as CTA.

We would like to thank NASA for support of this
  project through {\sl Fermi} Guest Investigator Program award 61101.
We gratefully acknowledge assistance from John Blondin on VH-1 issues.

\bibliographystyle{apj}
\bibliography{apj-jour,himodels}

\end{document}